\documentclass[prd,aps,floatfix,onecolumn,nofootinbib,amsmath,amssymb]{revtex4-2}

\usepackage[utf8]{inputenc}
\usepackage[T1]{fontenc}
\usepackage{lmodern}
\usepackage{bm}
\usepackage{amsmath}
\usepackage{hyperref}
\usepackage{graphicx}

\hypersetup{
    colorlinks=true,
    linkcolor=blue,
    citecolor=blue,
    urlcolor=blue,
}

\begin{document}

\title{Studying spherical collapse and its implications in the Eddington-inspired Born-Infeld gravity theory}

\author{Vel\'asquez-Toribio A.M.}
\email{alan.toribio@ufes.br}
\affiliation{Center for Astrophysics and Cosmology of the Esp\'{\i}rito Santo Federal University (UFES), Brazil.}

\date{\today}


\begin{abstract}
We investigate spherical collapse in Eddington-inspired Born--Infeld (EiBI) gravity in the subhorizon, pressureless, and quasi-static regime, emphasizing the matter-gradient correction that appears in the weak-field limit of the theory. Starting from the nonlinear continuity and Euler equations, we derive the evolution equation for the density contrast and show that the EiBI contribution depends explicitly on spatial derivatives of the matter density. This feature makes the ideal discontinuous top-hat construction ill-defined, since gradient terms become singular at the boundary, and requires a regularized overdensity profile together with a coarse-graining prescription. We adopt an effective physical-gradient closure for the EiBI source term and compare two matched initial configurations: a regularized Tanh profile and a peak-based profile, calibrated to share the same characteristic radius and cumulative mass proxy. Within this framework, we compute the linear collapse threshold $\delta_c(z_{\rm coll})$, the turnaround overdensity $\delta_t(z_{\rm coll})$, the turnaround radius $R_t(z_{\rm coll})$, and the virial overdensity $\Delta_{\rm vir}(z_{\rm coll})$. Relative to the $\Lambda$CDM reference case, the EiBI correction lowers $\delta_c$, enhances both $\delta_t$ and $\Delta_{\rm vir}$, and produces a more modest reduction of $R_t$, with deviations increasing with the dimensionless coupling $\hat\kappa_{\rm BI}$ over the range considered. The nonlinear overdensity observables show the strongest response to the EiBI correction and retain a residual dependence on the internal shape of the matched profile, whereas the turnaround radius is comparatively less affected. These results identify spherical collapse as a sensitive probe of EiBI matter-gradient couplings and motivate applications to halo statistics and nonlinear structure formation.
\end{abstract} 
\maketitle

\textbf{}\section{Introduction}
\label{sec:intro}

The observational establishment that the cosmic expansion is currently accelerating has brought renewed
attention to the physical origin of this phenomenon and to the fundamental status of general relativity
(GR) on cosmological scales. In the standard paradigm, the late-time acceleration is attributed to a
dark-energy sector with sufficiently negative effective pressure, ranging from a cosmological constant to
dynamical components with a time-dependent equation of state. In parallel, an extensive program has
pursued modifications of gravity as an alternative route, including scalar--tensor theories,
higher-curvature extensions such as metric and Palatini $f(R)$ models, and broader effective-field-theory
descriptions of departures from GR \cite{Peebles2003,BeltranJimenez2018,Perivolaropoulos2022,Nojiri2011,Nojiri2017,Heisenberg2019}. 
While many of these scenarios can be engineered to match the measured
background expansion history with high accuracy, they generally predict different patterns for the growth
of structure and for the nonlinear dynamics of gravitational collapse. This makes controlled studies of
structure formation a central avenue for distinguishing between dark-energy physics and genuine
modifications of gravity.

Eddington-inspired Born--Infeld (EiBI) gravity provides a particularly instructive realization of this
idea. Formulated in the Palatini approach, EiBI introduces a determinantal Born--Infeld structure that
interpolates between an Eddington-like affine action and the Einstein--Hilbert response in the presence of
matter \cite{Eddington1924, BornInfeld1934, BanadosFerreira2010, Avelino2012PRD_EiBI, AvelinoFerreira2012PRD, ScargillBanadosFerreira2012, PaniCardosoDelsate2011, DelsateSteinhoff2012, Makarenko2014}. A defining feature is that EiBI reduces exactly to GR (with an effective cosmological constant) in
vacuum, so that deviations from $\Lambda$CDM are not sourced by new long-range propagating degrees of
freedom, but are activated by local matter through an algebraic mapping between the physical metric and an
auxiliary metric compatible with the independent connection. In the weak-field, nonrelativistic regime
relevant for structure formation, this property manifests itself as a modified Poisson equation in which
the leading correction depends on spatial derivatives of the matter density. As a consequence, EiBI does
not merely rescale Newton's constant, but induces a profile-dependent force that can modify collapse,
equilibrium, and boundary-layer dynamics in a way that is absent in GR. This matter-gradient sensitivity
has been widely discussed in astrophysical contexts and is a key reason why EiBI remains an attractive and
nontrivial alternative-gravity framework to confront with data.

A standard and robust entry point into the nonlinear regime of structure formation is the spherical
collapse model. In GR, the pressureless top-hat idealization reduces the collapse dynamics to a single
degree of freedom and yields nearly universal benchmarks such as the linear collapse threshold
$\delta_c(z_{\rm coll})$ and the virial overdensity $\Delta_{\rm vir}(z)$ once the background cosmology is
specified \cite{GunnGott1972, PressSchechter1974, Bond1991ExcursionSet, Peebles1980LSS, LaceyCole1993, EkeColeFrenk1996, BryanNorman1998}. These quantities underpin widely used semi-analytic approaches, including halo mass functions,
bias models, and phenomenological mappings between linear and nonlinear statistics. For modified gravity,
the same formalism becomes a well-motivated diagnostic tool: deviations in $\delta_c$, turnaround
quantities, and virial overdensities propagate into halo abundances and clustering, thereby offering a
pathway to observational tests even when the background expansion is nearly degenerate with $\Lambda$CDM 
\cite{ChevallierPolarski2001CPL, Linder2003CPL, MotaVanDeBruck2004, PaceWaizmannBartelmann2010, PaceMeyerBartelmann2017Implementation, AbramoBatistaLiberatoRosenfeld2007, CreminelliEtAl2010ZeroCsCollapse, BasseBjaeldeWong2011ArbitraryCs, SefusattiVernizzi2011ClusteringQuintessence, WintergerstPettorino2010CoupledDEcollapse, TarrantVanDeBruckCopelandGreen2012CoupledQuintessenceHMF, BatistaDeOliveiraAbramo2023NonTopHatDE, DvaliGabadadzePorrati2000, Deffayet2001DGP, SchmidtHuLima2010DGPcollapse, KoyamaSilva2007DGPNonlinear, Vainshtein1972, BabichevDeffayet2013VainshteinReview, BarreiraLiBaughPascoli2013GalileonSC, HuSawicki2007, KhouryWeltman2004PRLChameleon, KhouryWeltman2004PRDChameleonCosmo, SchmidtLimaOyaizuHu2009FRHaloStats, LiHu2011ChameleonHaloModeling, LiEfstathiou2012ExtendedExcursionSetChameleon, KoppApplebyAchitouvWeller2013FRcollapse, LombriserLiKoyamaZhao2013ChameleonHMF, LombriserKoyamaLi2014HaloModellingChameleon, CataneoEtAl2016FRClusterAbundance, TaddeiCatenaPietroni2014SymmetronSC, BraxValageas2013ScreeningPowerSpectrum, BraxEtAl2010DilatonDamourPolyakov, BraxValageas2014KmouflageLSS,ShamLinLeung2014}.

In EiBI gravity, however, the standard top-hat reduction is not automatically applicable. Because the
weak-field modification depends explicitly on $\nabla\rho_m$ and $\nabla^2\rho_m$, a discontinuous density
profile generates distributional contributions at the boundary and the collapse dynamics is not uniquely
defined without an explicit coarse-graining prescription. The spherical-collapse problem in EiBI is
therefore intrinsically tied to the smoothing scale that regularizes boundary gradients, and collapse
observables are expected to acquire a controlled dependence on the adopted regularization 
\cite{PaniDelsateCardoso2012EiBIPhenomenology, DuChenGao2014EiBILSS, TavakoliEscamillaRiveraFabris2016EiBIcollapse, Lopes2018, Kopp2013, BBKS1986, JimenezHeisenbergKoivistoPekar2021}. A systematic
assessment of this effect - and of its implications for halo formation benchmarks - is still comparatively
underdeveloped relative to other modified-gravity scenarios, despite the distinctive character of EiBI
matter couplings.

In this work we develop the spherical-collapse formalism in EiBI gravity in the subhorizon, pressureless, and quasi-static regime. Starting from the exact nonlinear continuity and Euler equations, we derive a master evolution equation for the density contrast in terms of the scale factor, where EiBI departures from GR enter through the weak-field correction to the Poisson equation. By implementing a regularized spherical overdensity that ensures finite gradient terms, we quantify the impact of the smoothing scale on collapse dynamics, distinguishing our method from previous literature \cite{SantosSantos2015VirialEBI,Lombriser2015,Shao2020,SSham2013}. We then compute the linear collapse threshold $\delta_c(z_{\rm coll})$, the turnaround epoch, the turnaround overdensity, the turnaround radius, and the virial overdensity as functions of collapse redshift and mass, and we discuss virialization diagnostics that track controlled departures from homology induced by the gradient-driven EiBI force. Although the present analysis is restricted to the collapse sector, this restriction is deliberate rather than limiting: in EiBI gravity, any prediction for halo abundances, mass functions, bias, or number counts must rest on a prior and internally consistent determination of the nonlinear collapse criteria, since the theory modifies not only the effective force law but also the profile dependence of the collapsing patch through matter-gradient terms. In this sense, the spherical-collapse problem is not a secondary ingredient but the necessary foundation for any later phenomenological application to halo statistics. The purpose of this first paper is therefore to establish that foundation in a controlled manner and to isolate the impact of EiBI corrections on the benchmark quantities that enter semi-analytic large-scale-structure modelling. The extension of these results to halo abundance, halo number counts, and related observational predictions will be pursued in a subsequent work.

The paper is organized as follows. In Section II we review the main aspects of Eddington-inspired Born-Infeld gravity relevant to the present analysis. In Section III we develop the spherical-collapse framework and introduce the corresponding turnaround quantities and their physical implications. In Section IV we discuss the virial overdensity and the departures from the standard top-hat picture. In Section V we present the numerical results, and in Section VI we summarize our main conclusions and outline possible directions for future work.


\section{The Eddington-Inspired Born-Infeld gravity}
\label{sec:eibi_basic_aspects}

In this section we collect the basic equations of Eddington-inspired Born-Infeld gravity that will be
needed in the subsequent analysis. Since our main goal is to study spherical collapse and halo-related
observables in the weak-field regime, we present only the ingredients that are required later: the
action, the algebraic relation between the physical and auxiliary metrics, the homogeneous cosmological
background, and the nonrelativistic limit relevant for structure formation. In this sense, the present
section is not intended as a general review of the theory, but rather as a compact and self-contained
summary of the formulation that underlies the collapse equations developed in the following sections.

We work in the Palatini formulation, in which the physical metric $g_{\mu\nu}$ and the affine
connection $\Gamma^{\alpha}{}_{\mu\nu}$ are treated as independent variables. We assume a torsion-free
connection, so that the Ricci tensor constructed from $\Gamma^{\alpha}{}_{\mu\nu}$ is symmetric. We
adopt units $c=1$ and denote Newton's constant by $G_{\rm N}$. The physical metric has determinant
$g\equiv\det(g_{\mu\nu})$, while the symmetric Ricci tensor built from the independent connection is
denoted by $R_{\mu\nu}(\Gamma)$. Matter fields are collectively represented by $\Psi$ and are
minimally coupled to $g_{\mu\nu}$ through the matter action $S_M[g,\Psi]$. In this framework, an
auxiliary metric arises naturally and provides the most convenient route to both the homogeneous
background equations and the weak-field limit that will later be employed in our discussion of
spherical collapse and halo formation. The EiBI action is written as \cite{BanadosFerreira2010}
\begin{eqnarray}
S[g,\Gamma,\Psi]
&=&
\frac{1}{8\pi G_{\rm N}\kappa_{\rm BI}}
\int d^4x
\left[
\sqrt{\left|\,g_{\mu\nu}+\kappa_{\rm BI}R_{\mu\nu}(\Gamma)\,\right|}
-
\lambda\,\sqrt{|g|}
\right]
+S_M[g,\Psi],
\label{eq:eibi_action}
\end{eqnarray}
where $|\cdot|$ denotes the determinant of the matrix inside the brackets, $\kappa_{\rm BI}$ is the Born--Infeld parameter, with dimensions of length squared, and $\lambda$ is a dimensionless constant. The parameter $\lambda$ determines the effective cosmological constant through
\begin{eqnarray}
\Lambda_{\rm eff}=\frac{\lambda-1}{\kappa_{\rm BI}}.
\label{eq:Lambda_eff}
\end{eqnarray}

Thus, even before specifying a matter source, the theory already contains the parameter combination
that controls the low-curvature vacuum limit.

On the other hand, the matter stress-energy tensor is defined in the standard way by varying the matter action with
respect to the inverse physical metric:
\begin{eqnarray}
T_{\mu\nu}
\equiv
-\frac{2}{\sqrt{|g|}}
\frac{\delta S_M}{\delta g^{\mu\nu}},
\qquad
T^\mu{}_{\nu}\equiv g^{\mu\alpha}T_{\alpha\nu},
\label{eq:Tmunu_def}
\end{eqnarray}
and minimal coupling implies the standard covariant conservation law with respect to the Levi-Civita
derivative associated with $g_{\mu\nu}$,
\begin{eqnarray}
\nabla^{(g)}_{\mu}T^{\mu\nu}=0.
\label{eq:conservation}
\end{eqnarray}
This conservation equation keeps the same form as in general relativity because matter couples only
to the physical metric and not directly to the independent connection.

Varying Eq.~\eqref{eq:eibi_action} with respect to $g_{\mu\nu}$ yields
\begin{eqnarray}
\frac{\sqrt{|g+\kappa_{\rm BI}R|}}{\sqrt{|g|}}
\left[(g+\kappa_{\rm BI}R)^{-1}\right]^{\mu\nu}
-\lambda\,g^{\mu\nu}
=
-8\pi G_{\rm N}\kappa_{\rm BI}\,T^{\mu\nu},
\label{eq:eibi_metric_eq}
\end{eqnarray}
where $(g+\kappa_{\rm BI}R)^{-1}$ denotes the matrix inverse of
$g_{\mu\nu}+\kappa_{\rm BI}R_{\mu\nu}(\Gamma)$. The independent connection equation is more conveniently
handled after introducing an auxiliary metric $q_{\mu\nu}$ defined by
\begin{eqnarray}
q_{\mu\nu}=g_{\mu\nu}+\kappa_{\rm BI}\,R_{\mu\nu}(\Gamma),
\label{eq:q_def}
\end{eqnarray}
so that the variation with respect to $\Gamma^{\alpha}{}_{\mu\nu}$ implies
\begin{eqnarray}
\nabla^{(\Gamma)}_{\alpha}\left(\sqrt{|q|}\,q^{\mu\nu}\right)=0,
\label{eq:q_compat}
\end{eqnarray}
and therefore $\Gamma^\alpha{}_{\mu\nu}$ is the Levi-Civita connection of $q_{\mu\nu}$:
\begin{eqnarray}
\Gamma^\alpha{}_{\mu\nu}
=
\frac{1}{2}\,q^{\alpha\sigma}
\left(\partial_\mu q_{\sigma\nu}+\partial_\nu q_{\sigma\mu}-\partial_\sigma q_{\mu\nu}\right),
\label{eq:Gamma_LC_q}
\end{eqnarray}
with $q\equiv\det(q_{\mu\nu})$. Combining the metric and connection equations then gives the algebraic
relation
\begin{eqnarray}
\sqrt{|q|}\,q^{\mu\nu}
=
\sqrt{|g|}
\left(\lambda\,g^{\mu\nu}-8\pi G_{\rm N}\kappa_{\rm BI}\,T^{\mu\nu}\right).
\label{eq:master_relation}
\end{eqnarray}
Equation~\eqref{eq:master_relation} provides the key algebraic link between the matter sources and the
auxiliary geometry. It is this relation that will allow us to derive both the homogeneous cosmological
background equations and the weak-field limit relevant for the collapse problem. In particular, instead
of introducing new propagating vacuum degrees of freedom, the EiBI modification is encoded in the way
matter deforms the relation between the physical and auxiliary metrics.

To connect the general field equations with the cosmological regime relevant for structure formation,
we now specialize to a spatially flat Friedmann-Lema\^{i}tre-Robertson-Walker background filled with a
perfect fluid. For the physical metric we take
\begin{eqnarray}
ds_g^2=-dt^2+a(t)^2\delta_{ij}dx^i dx^j,
\qquad
H\equiv\frac{\dot a}{a},
\label{eq:flrw_g}
\end{eqnarray}
while for the matter source we assume
\begin{eqnarray}
T^\mu{}_{\nu}=\mathrm{diag}(-\rho,p,p,p).
\label{eq:perfect_fluid}
\end{eqnarray}
Under the same symmetry assumptions, the auxiliary metric may be parametrised as
\begin{eqnarray}
ds_q^2=-U(t)\,dt^2+a(t)^2V(t)\delta_{ij}dx^i dx^j.
\label{eq:flrw_q}
\end{eqnarray}

It is convenient to absorb the effective cosmological constant into total density and pressure variables,
defined by
\begin{eqnarray}
\rho_T=\rho+\Lambda_{\rm eff},
\qquad
P_T=p-\Lambda_{\rm eff}.
\label{eq:rhoT_PT}
\end{eqnarray}
At the same time, in order to keep the gravitational normalization explicit and avoid any ambiguity in
the dimensions of the Born-Infeld parameter, we introduce the rescaled combination
\begin{eqnarray}
\bar{\kappa}_{\rm BI}\equiv 8\pi G_{\rm N}\kappa_{\rm BI}.
\label{eq:kappa_bar}
\end{eqnarray}
With this notation, the products $\bar{\kappa}_{\rm BI}\rho_T$ and
$\bar{\kappa}_{\rm BI}P_T$ are dimensionless, and the homogeneous algebraic relations may be written
in a compact form. For a perfect-fluid FLRW source, Eq.~\eqref{eq:master_relation} can be solved
diagonally, yielding
\begin{eqnarray}
U=\frac{D}{1+\bar{\kappa}_{\rm BI}\rho_T},
\qquad
V=\frac{D}{1-\bar{\kappa}_{\rm BI}P_T},
\label{eq:UV}
\end{eqnarray}
where
\begin{eqnarray}
D\equiv
\sqrt{
\left(1+\bar{\kappa}_{\rm BI}\rho_T\right)
\left(1-\bar{\kappa}_{\rm BI}P_T\right)^3
}.
\label{eq:D}
\end{eqnarray}
The quantity $D$ is therefore not an independent dynamical variable, but a convenient algebraic
abbreviation that appears when one solves the relation between the physical and auxiliary metrics in
the homogeneous background. It encodes the determinant structure inherited from the EiBI field
equations and allows the time and space components of the auxiliary metric to be written in a compact
way. The background continuity equation still follows directly from Eq.~\eqref{eq:conservation}:
\begin{eqnarray}
\dot\rho=-3H(\rho+p).
\label{eq:continuity}
\end{eqnarray}
Therefore, when $|\bar{\kappa}_{\rm BI}\rho_T|\ll 1$ and
$|\bar{\kappa}_{\rm BI}P_T|\ll 1$, one has $U\simeq 1$ and $V\simeq 1$, so that the homogeneous
cosmological evolution approaches that of $\Lambda$CDM with $\Lambda=\Lambda_{\rm eff}$. Notice,
however, that this recovery of the standard background does not imply the complete absence of
modifications in the presence of inhomogeneities, since the theory remains sensitive to matter
gradients once one moves to the weak-field perturbed regime.

We now turn to the nonrelativistic limit relevant for structure formation. In this regime, the EiBI
field equations lead to a modified Poisson equation for the Newtonian potential $\Phi$ in physical
coordinates $\mathbf{r}$ \cite{BanadosFerreira2010}:
\begin{eqnarray}
\nabla_{\mathbf{r}}^2\Phi
=
4\pi G_{\rm N}\rho_m
+\frac{\kappa_{\rm BI}}{4}\nabla_{\mathbf{r}}^2\rho_m,
\label{eq:poisson_eibi_physical}
\end{eqnarray}
where $\rho_m$ is the nonrelativistic matter density. This expression makes explicit that the leading
EiBI correction is not simply a rescaling of Newton's constant. Instead, it introduces a contribution
that depends directly on spatial derivatives of the density field, and therefore on the internal
structure of the matter distribution.

Introducing the shifted potential
\begin{eqnarray}
\Phi'=\Phi-\frac{\kappa_{\rm BI}}{4}\rho_m,
\label{eq:phi_prime_def}
\end{eqnarray}
one obtains the equivalent Poisson equation
\begin{eqnarray}
\nabla_{\mathbf{r}}^2\Phi'=4\pi G_{\rm N}\rho_m,
\label{eq:phi_prime_poisson}
\end{eqnarray}
so that the matter acceleration can be written as
\begin{eqnarray}
\mathbf{a}_m=-\nabla_{\mathbf{r}}\Phi'
-\frac{\kappa_{\rm BI}}{4}\nabla_{\mathbf{r}}\rho_m.
\label{eq:acceleration_eibi_physical}
\end{eqnarray}
This form is especially useful for the applications considered later, because it separates the standard
Newtonian contribution from the genuine EiBI correction. The latter is proportional to the density
gradient and therefore vanishes for strictly uniform matter distributions. As a result, in contrast
with the standard top-hat treatment in general relativity, the collapse dynamics in EiBI gravity is
sensitive not only to the overdensity amplitude but also to the detailed radial structure of the
profile. This point is central for the spherical-collapse problem developed in the next section, where
the necessity of a regularized overdensity profile follows directly from the gradient-dependent nature
of the EiBI force.

\section{Spherical collapse in EiBI gravity}
\label{sec:spherical_collapse_eibi}

In this section we derive the nonlinear evolution equation for the matter density contrast in Eddington-inspired Born--Infeld gravity and introduce the turnaround quantities used in the subsequent analysis. We restrict attention to pressureless matter on subhorizon scales and work within the Newtonian and quasi-static approximations. The EiBI corrections employed below follow from the standard nonrelativistic weak-field limit of the theory \cite{BanadosFerreira2010}. In this regime, the modified force law depends not only on the density itself but also on its spatial derivatives. Consequently, unlike in general relativity, the collapse dynamics is not specified solely by the initial overdensity amplitude, but also by the smoothing prescription used to render density gradients finite and physically meaningful within the fluid description.

We work with both physical and comoving coordinates. The physical position $\mathbf r$ and the comoving position $\mathbf x$ are related by
\begin{eqnarray}
\mathbf r=a(t)\mathbf x,
\qquad
\nabla_{\mathbf r}=a^{-1}\nabla_{\mathbf x},
\qquad
\nabla_{\mathbf r}^{2}=a^{-2}\nabla_{\mathbf x}^{2},
\label{eq:physical_comoving_derivatives}
\end{eqnarray}
where $a(t)$ is the scale factor and $H\equiv\dot a/a$ is the Hubble rate. The matter density is decomposed as
\begin{eqnarray}
\rho_m(\mathbf{x},t)
=
\bar\rho_m(t)\,[1+\delta(\mathbf{x},t)],
\label{eq:rho_split_eibi}
\end{eqnarray}
where $\bar\rho_m(t)$ is the homogeneous background density and $\delta(\mathbf{x},t)$ is the density contrast. The peculiar velocity field is denoted by $\mathbf v(\mathbf x,t)$, and we define its comoving divergence by
\begin{eqnarray}
\theta(\mathbf{x},t)
\equiv
\nabla_{\mathbf x}\cdot\mathbf v .
\label{eq:theta_def_eibi}
\end{eqnarray}

Since the weak-field EiBI equations contain terms proportional to $\nabla\rho_m$ and $\nabla^2\rho_m$, the density entering the effective fluid equations must be understood as a coarse-grained field rather than as the microscopic particle density. We therefore define
\begin{eqnarray}
\rho_m(\mathbf{x},t)
\equiv
\int d^3x'\,
W_\ell(|\mathbf{x}-\mathbf{x}'|)
\,\rho_{\rm mic}(\mathbf{x}',t),
\qquad
\int d^3x\,W_\ell(|\mathbf{x}|)=1,
\label{eq:smoothing_def}
\end{eqnarray}
where $\rho_{\rm mic}$ is the microscopic density, $W_\ell$ is a normalized smoothing kernel, and $\ell$ denotes a coarse-graining scale. The density contrast introduced in Eq.~\eqref{eq:rho_split_eibi} inherits the same smoothing prescription. In practice, the dependence on the smoothing procedure will be encoded through a profile factor that measures the strength of the density-gradient correction for a given regularized profile.

For pressureless matter, the Newtonian continuity and Euler equations in an expanding background read \cite{Peebles1980LSS,LaceyCole1993,EkeColeFrenk1996,BryanNorman1998}:

\begin{eqnarray}
\dot\delta
+
\frac{1}{a}
\nabla_{\mathbf x}\cdot
\left[
(1+\delta)\mathbf v
\right]
=
0,
\label{eq:continuity_eibi}
\\
\dot{\mathbf v}
+
H\mathbf v
+
\frac{1}{a}
\left(
\mathbf v\cdot\nabla_{\mathbf x}
\right)\mathbf v
=
-\frac{1}{a}
\nabla_{\mathbf x}\Phi ,
\label{eq:euler_eibi}
\end{eqnarray}
where $\Phi(\mathbf x,t)$ is the peculiar gravitational potential. Assuming negligible primordial vorticity, the standard spherical-collapse closure amounts to neglecting shear and maintaining an irrotational flow. Taking the divergence of Eq.~\eqref{eq:euler_eibi}, eliminating $\theta$ with the aid of Eq.~\eqref{eq:continuity_eibi}, and retaining the exact nonlinear terms, one obtains
\begin{eqnarray}
\ddot\delta
+
2H\dot\delta
-
\frac{4}{3}
\frac{\dot\delta^2}{1+\delta}
=
\frac{1+\delta}{a^2}
\nabla_{\mathbf x}^{2}\Phi .
\label{eq:delta_master_t_eibi}
\end{eqnarray}
This relation is purely kinematical and does not yet depend on the gravitational model. Rewriting time derivatives in terms of derivatives with respect to the scale factor, using $d/dt=aH\,d/da$ and denoting $d/da$ by a prime, we find
\begin{eqnarray}
\delta''
+
\left(
\frac{3}{a}
+
\frac{H'}{H}
\right)\delta'
-
\frac{4}{3}
\frac{\delta'^2}{1+\delta}
=
\frac{1+\delta}{H^2a^4}
\nabla_{\mathbf x}^{2}\Phi .
\label{eq:delta_master_a_eibi}
\end{eqnarray}
At this stage, all model dependence is contained in the relation between $\nabla_{\mathbf x}^{2}\Phi$ and the density field.

In the weak-field nonrelativistic limit of EiBI gravity, the modified Poisson equation is most naturally written in physical coordinates as
\begin{eqnarray}
\nabla_{\mathbf r}^{2}\Phi
=
4\pi G_{\rm N}\rho_m
+
\frac{\kappa_{\rm BI}}{4}
\nabla_{\mathbf r}^{2}\rho_m .
\label{eq:poisson_eibi_physical}
\end{eqnarray}
After subtracting the homogeneous background contribution, using Eq.~\eqref{eq:rho_split_eibi}, and converting to comoving variables, one obtains
\begin{eqnarray}
\nabla_{\mathbf x}^{2}\Phi
=
4\pi G_{\rm N}a^2\bar\rho_m\,\delta
+
\frac{\kappa_{\rm BI}}{4}
\bar\rho_m
\nabla_{\mathbf x}^{2}\delta .
\label{eq:poisson_eibi_comoving}
\end{eqnarray}
The EiBI term in Eq.~\eqref{eq:poisson_eibi_comoving} does not carry an additional factor $a^2$. The factor $a^{-2}$ coming from $\nabla_{\mathbf r}^{2}=a^{-2}\nabla_{\mathbf x}^{2}$ is exactly compensated when Eq.~\eqref{eq:poisson_eibi_physical} is multiplied by $a^2$ to obtain the comoving Poisson equation. This observation is useful for avoiding double counting of scale-factor powers.

Substituting Eq.~\eqref{eq:poisson_eibi_comoving} into Eq.~\eqref{eq:delta_master_a_eibi} gives the exact coarse-grained nonlinear equation in terms of the comoving Laplacian,
\begin{eqnarray}
\delta''
+
\left(
\frac{3}{a}
+
\frac{H'}{H}
\right)\delta'
-
\frac{4}{3}
\frac{\delta'^2}{1+\delta}
&=&
\frac{3}{2}
\frac{\Omega_m(a)}{a^2}
(1+\delta)\delta
\nonumber\\
&&
+
\frac{\kappa_{\rm BI}\bar\rho_m}{4H^2a^4}
(1+\delta)
\nabla_{\mathbf x}^{2}\delta ,
\label{eq:delta_master_eibi_comoving_laplacian}
\end{eqnarray}
where
\begin{eqnarray}
\Omega_m(a)
\equiv
\frac{8\pi G_{\rm N}\bar\rho_m}{3H^2}.
\label{eq:Omega_m_def_eibi}
\end{eqnarray}
Equivalently,
\begin{eqnarray}
\delta''
+
\left(
\frac{3}{a}
+
\frac{H'}{H}
\right)\delta'
-
\frac{4}{3}
\frac{\delta'^2}{1+\delta}
&=&
\frac{3}{2}
\frac{\Omega_m(a)}{a^2}
(1+\delta)\delta
\nonumber\\
&&
+
\frac{3\kappa_{\rm BI}}{32\pi G_{\rm N}}
\frac{\Omega_m(a)}{a^4}
(1+\delta)
\nabla_{\mathbf x}^{2}\delta .
\label{eq:delta_master_eibi_comoving}
\end{eqnarray}
This form makes clear that, if the Laplacian is explicitly taken with respect to comoving coordinates, the EiBI source is accompanied by the combination $a^{-4}\nabla_{\mathbf x}^{2}\delta$.

The same term may also be written in terms of the physical Laplacian. Since
\begin{eqnarray}
\nabla_{\mathbf x}^{2}\delta
=
a^2\nabla_{\mathbf r}^{2}\delta ,
\label{eq:laplacian_physical_comoving_relation}
\end{eqnarray}
the EiBI contribution in Eq.~\eqref{eq:delta_master_eibi_comoving} can equivalently be expressed as
\begin{eqnarray}
\frac{\Omega_m(a)}{a^4}
\nabla_{\mathbf x}^{2}\delta
=
\frac{\Omega_m(a)}{a^2}
\nabla_{\mathbf r}^{2}\delta .
\label{eq:eibi_laplacian_equivalence}
\end{eqnarray}
Thus, the exact differential equation is independent of the coordinate convention. The apparent power of $a$ changes only because the Laplacian has been written with respect to different spatial coordinates.

The distinction becomes physically relevant when the Laplacian is replaced by an effective coarse-graining scale. In the numerical implementation used below, we adopt a physical-gradient closure for the EiBI correction,
\begin{eqnarray}
\nabla_{\mathbf r}^{2}\delta
\simeq
-
k_{\rm phys}^{2}\,{\cal P}\,\delta ,
\label{eq:laplacian_physical_closure}
\end{eqnarray}
where $k_{\rm phys}$ is an effective physical smoothing wavenumber and ${\cal P}$ is a dimensionless profile factor encoding the dependence on the regularized density profile. With this prescription, Eq.~\eqref{eq:delta_master_eibi_comoving} becomes
\begin{eqnarray}
\delta''
+
\left(
\frac{3}{a}
+
\frac{H'}{H}
\right)\delta'
-
\frac{4}{3}
\frac{\delta'^2}{1+\delta}
&=&
\frac{3}{2}
\frac{\Omega_m(a)}{a^2}
(1+\delta)\delta
\nonumber\\
&&
-
\hat\kappa_{\rm BI}
\frac{\Omega_m(a)}{a^2}
\left(
\frac{k_{\rm phys}}{H_0/c}
\right)^2
{\cal P}
(1+\delta)\delta .
\label{eq:delta_master_eibi_effective}
\end{eqnarray}
Here $\hat\kappa_{\rm BI}$ denotes the dimensionless EiBI coupling used in the numerical implementation. Equation~\eqref{eq:delta_master_eibi_effective} is the effective equation solved in the numerical analysis. It corresponds to a physical coarse-graining prescription for the density-gradient correction. If instead one fixes a comoving/Lagrangian smoothing scale and writes
\begin{eqnarray}
\nabla_{\mathbf x}^{2}\delta
\simeq
-
k_{\rm com}^{2}\,{\cal P}\,\delta ,
\end{eqnarray}
then the corresponding effective EiBI source would be proportional to $\Omega_m(a)/a^4$. The two prescriptions are not merely different notations: they encode different assumptions about the scale at which the density-gradient term is coarse-grained. In the present work we use Eq.~\eqref{eq:laplacian_physical_closure}, which is the prescription implemented in the code and allows the EiBI coupling to be compared directly through the dimensionless parameter $\hat\kappa_{\rm BI}$.

Under spherical symmetry, the density contrast depends only on the radial coordinate and on the scale factor. If the radial coordinate is comoving, $r_x=|\mathbf x|$, then
\begin{eqnarray}
\nabla_{\mathbf x}^{2}\delta(r_x,a)
=
\frac{1}{r_x^2}
\frac{\partial}{\partial r_x}
\left(
r_x^2
\frac{\partial\delta}{\partial r_x}
\right).
\label{eq:laplacian_spherical_comoving}
\end{eqnarray}
If the radial coordinate is physical, $r=|\mathbf r|=a r_x$, then
\begin{eqnarray}
\nabla_{\mathbf r}^{2}\delta(r,a)
=
\frac{1}{r^2}
\frac{\partial}{\partial r}
\left(
r^2
\frac{\partial\delta}{\partial r}
\right).
\label{eq:laplacian_spherical_physical}
\end{eqnarray}
Equations~\eqref{eq:laplacian_spherical_comoving} and \eqref{eq:laplacian_spherical_physical} are related by Eq.~\eqref{eq:laplacian_physical_comoving_relation}. In the effective treatment adopted here, the gradient contribution is evaluated through the physical closure \eqref{eq:laplacian_physical_closure}.

The dependence on spatial derivatives is the main structural difference relative to the standard top-hat treatment in general relativity. An ideal top-hat profile,
\begin{eqnarray}
\delta(r,a_i)
=
\delta_0\Theta(r_b-r),
\end{eqnarray}
is discontinuous at the boundary and generates distributional contributions in both the first derivative and the Laplacian of the density contrast. The EiBI correction is then ill-defined unless a regularization prescription is specified. In this sense, smooth coarse graining is not merely a technical convenience, but part of the definition of the collapse problem itself.

A simple regularized profile is
\begin{eqnarray}
\delta(r,a_i)
=
\frac{\delta_0}{2}
\left[
1-
\tanh\left(
\frac{r/r_b-1}{s}
\right)
\right],
\label{eq:tanh_profile}
\end{eqnarray}
where $\delta_0$ is the central amplitude, $r_b$ sets the characteristic size of the perturbation, and $s>0$ controls the width of the transition layer. This form ensures finite first and second radial derivatives. In practice, $s$ parametrizes the smoothing scale, and its value fixes the strength of the EiBI gradient contribution through the profile factor ${\cal P}$.

In addition to the regularized Tanh profile, we also consider a second class of initial conditions constructed from peaks of a Gaussian random field, which we refer to as the peak-based profile. Unlike the Tanh ansatz, this profile is not introduced merely as a phenomenological smoothing prescription. Instead, it is defined as the mean density contrast around a selected peak and therefore admits a direct statistical interpretation.

Let $\delta(z_i,\mathbf{x},R)$ denote the initial density contrast at redshift $z_i$, smoothed on the scale $R$. The peak-based profile is defined through the conditional average \cite{BBKS1986}
\begin{eqnarray}
\delta_i(r,R)
=
\left\langle
\delta(z_i,\mathbf{x},R)
\,\big|\,
{\rm peak},\nu
\right\rangle ,
\label{eq:phy_profile_definition}
\end{eqnarray}
where $r$ is the distance from the center of the peak and $\nu$ is the dimensionless peak height, defined by
\begin{eqnarray}
\nu
=
\frac{\delta_{i,0}}{\sigma_i(R)},
\label{eq:phy_nu_definition}
\end{eqnarray}
with $\delta_{i,0}$ the central overdensity amplitude and $\sigma_i(R)$ the rms fluctuation of the smoothed linear density field on scale $R$.

Assuming spherical symmetry, the mean profile can be written in Fourier space as
\begin{eqnarray}
\delta_i(r,R)
=
\frac{2}{\pi}
\int_{0}^{\infty}
dk\,
k^2
\delta_0(k,R)
\frac{\sin(kr)}{kr}
T(k),
\label{eq:phy_profile_fourier}
\end{eqnarray}
where $k$ is the wavenumber, $T(k)$ is the matter transfer function, and $\sin(kr)/(kr)$ is the spherical Bessel kernel associated with the inverse transform of a spherically symmetric configuration. This is the standard representation adopted for mean peak profiles in Refs.~\cite{Lopes2018,Kopp2013,BBKS1986}.

The primordial shape function is parameterized as
\begin{eqnarray}
\delta_0(k,R)
=
\delta_{i,0}
\frac{1}{4\pi (n_s+5)R^3}
e^{-k^2R^2}
(kR)^{n_s}
F(\nu,n_s,k,R),
\label{eq:phy_delta0_definition}
\end{eqnarray}
where $n_s$ is the scalar spectral index. The function $F(\nu,n_s,k,R)$ encodes the peak constraint and the statistical corrections associated with conditioning on a local maximum. Equivalently, it specifies how the average profile around a selected peak differs from a purely phenomenological regularization.

The distinction between the Tanh and peak-based profiles is particularly relevant here. In EiBI gravity the evolution depends explicitly on spatial derivatives of the density field, so two perturbations with the same characteristic mass scale need not evolve identically if their radial structures differ. Comparing these two classes of initial conditions therefore provides a direct way to assess the sensitivity of the collapse threshold and turnaround observables to the internal shape of the overdensity.

We now turn to the definition of turnaround. For a shell labeled by comoving radius $r_s(a)$, the corresponding physical radius is
\begin{eqnarray}
R(a)
\equiv
a\,r_s(a),
\label{eq:R_def_shell}
\end{eqnarray}
and its physical radial velocity is
\begin{eqnarray}
\dot R
=
HR+v_r,
\qquad
v_r
\equiv
a\,\dot r_s .
\label{eq:Rdot_def_shell}
\end{eqnarray}
Turnaround occurs at the scale factor $a=a_t$ for which the shell momentarily stops expanding in physical coordinates,
\begin{eqnarray}
\dot R(a_t)=0,
\end{eqnarray}
or equivalently,
\begin{eqnarray}
v_r(a_t)
=
-H(a_t)R(a_t).
\label{eq:turnaround_condition_velocity}
\end{eqnarray}

Under spherical symmetry, the velocity divergence is
\begin{eqnarray}
\theta(r,a)
=
\frac{1}{r^2}
\frac{\partial}{\partial r}
\left(
r^2v_r
\right).
\label{eq:theta_spherical}
\end{eqnarray}
If the interior flow is approximately homologous, then $v_r\propto r$, so that $v_r=(\theta/3)r$. In this limit, the turnaround condition at the outer boundary may also be written as
\begin{eqnarray}
\theta(a_t)
=
-3a_tH(a_t).
\label{eq:turnaround_theta}
\end{eqnarray}
This provides a useful bulk diagnostic when the regularized profile remains close to top-hat-like in its interior. In the numerical analysis, however, turnaround will be identified directly from the shell condition $\dot R=0$, since this remains unambiguous for smooth profiles with finite gradients.

The turnaround radius and overdensity can be expressed in terms of a dimensionless radius variable. For a uniform interior region of conserved mass $M$ and volume-averaged density contrast $\delta_{\rm th}(a)$, mass conservation implies
\begin{eqnarray}
1+\delta_{\rm th}(a)
=
\left(
\frac{a}{a_i}
\right)^3
\left(
\frac{R_i}{R(a)}
\right)^3
[1+\delta_{\rm th}(a_i)].
\label{eq:delta_R_relation}
\end{eqnarray}
Introducing
\begin{eqnarray}
y(a)
\equiv
\frac{R(a)}{aR_i/a_i},
\label{eq:y_def}
\end{eqnarray}
one finds
\begin{eqnarray}
1+\delta_{\rm th}(a)
=
\frac{1+\delta_{\rm th}(a_i)}{y(a)^3},
\label{eq:delta_y_relation}
\end{eqnarray}
while the turnaround condition becomes
\begin{eqnarray}
y'(a_t)
=
-\frac{y(a_t)}{a_t}.
\label{eq:y_turnaround_a}
\end{eqnarray}
The corresponding turnaround radius and turnaround overdensity are then
\begin{eqnarray}
R_t
\equiv
R(a_t)
=
\frac{a_t}{a_i}R_i\,y(a_t),
\label{eq:R_turnaround_def}
\\
\delta_t
\equiv
\delta_{\rm th}(a_t)
=
\frac{1+\delta_{\rm th}(a_i)}{y(a_t)^3}
-
1.
\label{eq:delta_turnaround_def}
\end{eqnarray}

In EiBI gravity, the quantities $(a_t,R_t,\delta_t)$ inherit their dependence on $\kappa_{\rm BI}$ through the gradient terms entering the effective source in Eq.~\eqref{eq:delta_master_eibi_effective}. Turnaround predictions are therefore necessarily tied to the adopted smoothing prescription, which fixes the structure of the transition layer and hence the magnitude of the effective Laplacian. In the numerical section we adopt the physical-gradient coarse-graining prescription of Eq.~\eqref{eq:laplacian_physical_closure}, compare regularized Tanh and peak-based profiles, and estimate the resulting profile dependence while keeping the enclosed mass fixed. The numerical integration of Eq.~\eqref{eq:delta_master_eibi_effective}, together with the turnaround diagnostics introduced above, is deferred to the next section.

\section{Virial overdensity from an effective virial prescription in EiBI gravity}
\label{sec:virial_eibi}

In the previous section we showed that, in EiBI gravity, the nonlinear evolution of the density contrast depends explicitly on spatial derivatives of the matter profile through the gradient correction entering the weak-field force law. This feature has a direct consequence for the late stages of collapse. The usual general-relativistic top-hat reduction to a single exactly homogeneous degree of freedom is no longer strictly applicable once the EiBI correction is retained, since an ideal discontinuous top-hat profile produces ill-defined boundary gradients. Therefore, the virialization stage in EiBI gravity must be treated with some care.

The purpose of this section is not to derive a universal virial theorem for arbitrary EiBI configurations. Rather, we construct an effective virial prescription adapted to regularized spherical collapse. This prescription is meant to provide a controlled and physically motivated estimate of the virial radius and of the corresponding virial overdensity once a smooth coarse-grained density profile has been specified. Consequently, the virial overdensity obtained in this way should not be interpreted as a universal function of redshift alone. It is instead an effective collapse observable that may depend on the EiBI coupling, on the enclosed mass, on the smoothing prescription, and on the structure of the transition layer.

This point is particularly important in EiBI gravity. Since the weak-field correction is sourced by density gradients, the virial balance can receive contributions from the bulk profile and from boundary terms. Thus, a purely imposed prescription such as $R_{\rm vir}=\eta R_t$ would hide part of the profile dependence that is characteristic of the theory. The derivation below should therefore be viewed as a useful effective framework, not as an exact replacement for a full dynamical treatment of the regularized density profile.

\subsection{Effective virial relation in the weak-field EiBI regime}

The virial relation is most naturally written in physical coordinates. In the nonrelativistic weak-field limit relevant for the present analysis, the matter acceleration may be written as
\begin{eqnarray}
\mathbf{a}
=
-\nabla_{\mathbf r}\Phi'
-\frac{\kappa_{\rm BI}}{4}\nabla_{\mathbf r}\rho ,
\label{eq:eibi_acceleration_virial}
\end{eqnarray}
where $\mathbf r$ is the physical position, $\rho$ is the physical matter density, and $\Phi'$ satisfies the standard Poisson equation for the shifted potential,
\begin{eqnarray}
\nabla_{\mathbf r}^{2}\Phi'
=
4\pi G_{\rm N}\rho .
\label{eq:shifted_poisson_virial}
\end{eqnarray}
The first term in Eq.~\eqref{eq:eibi_acceleration_virial} is the usual Newtonian contribution written in terms of the shifted potential, while the second term is the EiBI correction associated with density gradients. The corresponding force density is
\begin{eqnarray}
\mathbf{f}
=
\rho\,\mathbf{a}
=
-\rho\nabla_{\mathbf r}\Phi'
-\frac{\kappa_{\rm BI}}{4}\rho\nabla_{\mathbf r}\rho .
\label{eq:force_density_virial}
\end{eqnarray}

We introduce the scalar moment of inertia of the collapsing region,
\begin{eqnarray}
I(t)
\equiv
\int_V \rho(\mathbf r,t)\,r^2\,dV ,
\label{eq:moment_inertia_def}
\end{eqnarray}
where $V$ denotes the physical volume occupied by the regularized overdensity. Applying the standard virial manipulation to the pressureless effective fluid, one obtains \cite{Peebles1980LSS}
\begin{eqnarray}
\frac{1}{2}\ddot I
=
2T+W_{\rm N}+W_{\rm EiBI},
\label{eq:virial_preliminary}
\end{eqnarray}
where
\begin{eqnarray}
T
\equiv
\frac{1}{2}\int_V \rho\,v^2\,dV
\label{eq:kinetic_energy_def}
\end{eqnarray}
is the kinetic energy associated with the peculiar motion, and
\begin{eqnarray}
W_{\rm N}
\equiv
\int_V
\rho\,\mathbf r\cdot
\left(-\nabla_{\mathbf r}\Phi'\right)\,dV
\label{eq:WN_def}
\end{eqnarray}
is the Newtonian gravitational virial contribution. The additional EiBI contribution is
\begin{eqnarray}
W_{\rm EiBI}
\equiv
-\frac{\kappa_{\rm BI}}{4}
\int_V
\rho\,\mathbf r\cdot\nabla_{\mathbf r}\rho\,dV .
\label{eq:WEiBI_initial}
\end{eqnarray}

Using
\begin{eqnarray}
\rho\nabla_{\mathbf r}\rho
=
\frac{1}{2}\nabla_{\mathbf r}(\rho^2),
\label{eq:rho_grad_identity}
\end{eqnarray}
we may rewrite Eq.~\eqref{eq:WEiBI_initial} as
\begin{eqnarray}
W_{\rm EiBI}
=
-\frac{\kappa_{\rm BI}}{8}
\int_V
\mathbf r\cdot\nabla_{\mathbf r}(\rho^2)\,dV .
\label{eq:WEiBI_rewrite}
\end{eqnarray}
The identity
\begin{eqnarray}
\nabla_{\mathbf r}\cdot(\rho^2\mathbf r)
=
3\rho^2+\mathbf r\cdot\nabla_{\mathbf r}(\rho^2)
\label{eq:divergence_identity}
\end{eqnarray}
then gives
\begin{eqnarray}
\int_V
\mathbf r\cdot\nabla_{\mathbf r}(\rho^2)\,dV
=
\oint_{\partial V}
\rho^2\,\mathbf r\cdot d\mathbf S
-
3\int_V \rho^2\,dV .
\label{eq:integration_by_parts_density}
\end{eqnarray}
Therefore,
\begin{eqnarray}
W_{\rm EiBI}
=
\frac{3\kappa_{\rm BI}}{8}
\int_V \rho^2\,dV
-
\frac{\kappa_{\rm BI}}{8}
\oint_{\partial V}
\rho^2\,\mathbf r\cdot d\mathbf S .
\label{eq:WEiBI_final}
\end{eqnarray}

Substituting Eq.~\eqref{eq:WEiBI_final} into Eq.~\eqref{eq:virial_preliminary}, we find
\begin{eqnarray}
\frac{1}{2}\ddot I
=
2T+W_{\rm N}
+
\frac{3\kappa_{\rm BI}}{8}
\int_V \rho^2\,dV
-
\frac{\kappa_{\rm BI}}{8}
\oint_{\partial V}
\rho^2\,\mathbf r\cdot d\mathbf S .
\label{eq:generalized_virial_eibi}
\end{eqnarray}
At an approximately stationary virial stage, for which $\ddot I\simeq0$, this gives the effective balance
\begin{eqnarray}
2T+W_{\rm N}
+
\frac{3\kappa_{\rm BI}}{8}
\int_V \rho^2\,dV
-
\frac{\kappa_{\rm BI}}{8}
\oint_{\partial V}
\rho^2\,\mathbf r\cdot d\mathbf S
\simeq0 .
\label{eq:virial_condition_surface}
\end{eqnarray}

It is useful to define
\begin{eqnarray}
U_{\rm EiBI}
\equiv
\frac{\kappa_{\rm BI}}{8}
\int_V \rho^2\,dV ,
\label{eq:Ueibi_def}
\end{eqnarray}
and
\begin{eqnarray}
\Sigma_{\rm EiBI}
\equiv
\frac{\kappa_{\rm BI}}{8}
\oint_{\partial V}
\rho^2\,\mathbf r\cdot d\mathbf S .
\label{eq:surface_term_def}
\end{eqnarray}
The effective virial condition can then be written as
\begin{eqnarray}
2T+W_{\rm N}+3U_{\rm EiBI}-\Sigma_{\rm EiBI}\simeq0 .
\label{eq:virial_condition_compact}
\end{eqnarray}

Equation~\eqref{eq:virial_condition_compact} should be interpreted as the virial balance associated with the weak-field EiBI force law for a regularized density configuration. Its main role is to display how the matter-gradient correction modifies the standard virial structure: it generates a bulk term proportional to $\int\rho^2dV$ and, in general, a surface contribution controlled by the density profile near the boundary. The presence of these terms is specific to EiBI gravity and is the reason why virial quantities in this framework are expected to retain a dependence on the adopted smoothing prescription.

\subsection{Recovery of the general-relativistic limit}

The standard general-relativistic virial relation is recovered in the limit
\begin{eqnarray}
\kappa_{\rm BI}\rightarrow0 .
\label{eq:kappa_to_zero}
\end{eqnarray}
In this case,
\begin{eqnarray}
U_{\rm EiBI}\rightarrow0,
\qquad
\Sigma_{\rm EiBI}\rightarrow0,
\end{eqnarray}
and Eq.~\eqref{eq:virial_condition_compact} reduces to
\begin{eqnarray}
2T+W_{\rm N}=0 .
\label{eq:virial_gr_limit}
\end{eqnarray}
This is the familiar virial relation used in the Newtonian and general-relativistic top-hat descriptions of pressureless collapse.

This limit is useful as a consistency check. It shows that the effective EiBI construction is continuously connected with the standard spherical-collapse picture. Deviations from the usual virial overdensity arise only through the matter-gradient coupling encoded in $\kappa_{\rm BI}$ and through the profile-dependent quantities $U_{\rm EiBI}$ and $\Sigma_{\rm EiBI}$.

\subsection{Energy balance between turnaround and virialization}

The effective virial balance may be combined with an approximate energy argument in order to estimate the virial radius. We define an effective energy functional for the regularized collapsing region as
\begin{eqnarray}
E
=
T+W_{\rm N}+U_{\rm EiBI}.
\label{eq:total_energy_eibi}
\end{eqnarray}
This expression should not be read as an exact conserved energy for an arbitrary evolving EiBI configuration. It is a working approximation for the coarse-grained spherical-collapse model. In particular, when the surface contribution $\Sigma_{\rm EiBI}$ is non-negligible, the energy-balance relation may receive boundary corrections. In the leading estimate used below, the surface term is either neglected or interpreted as part of the systematic uncertainty associated with the effective virial prescription.

At turnaround, the peculiar kinetic energy is approximately zero, $T_t\simeq0$, and therefore
\begin{eqnarray}
E_t
\simeq
W_{{\rm N},t}+U_{{\rm EiBI},t}.
\label{eq:energy_turnaround}
\end{eqnarray}
At virialization,
\begin{eqnarray}
E_{\rm vir}
=
T_{\rm vir}
+
W_{{\rm N},{\rm vir}}
+
U_{{\rm EiBI},{\rm vir}} .
\label{eq:energy_vir}
\end{eqnarray}
Using Eq.~\eqref{eq:virial_condition_compact} while neglecting $\Sigma_{\rm EiBI}$ at leading order gives
\begin{eqnarray}
2T_{\rm vir}
+
W_{{\rm N},{\rm vir}}
+
3U_{{\rm EiBI},{\rm vir}}
\simeq0 ,
\label{eq:virial_no_surface}
\end{eqnarray}
and hence
\begin{eqnarray}
T_{\rm vir}
\simeq
-\frac{1}{2}
\left(
W_{{\rm N},{\rm vir}}
+
3U_{{\rm EiBI},{\rm vir}}
\right).
\label{eq:Tvir_from_virial}
\end{eqnarray}
The effective energy at virialization is then
\begin{eqnarray}
E_{\rm vir}
\simeq
\frac{1}{2}
W_{{\rm N},{\rm vir}}
-
\frac{1}{2}
U_{{\rm EiBI},{\rm vir}} .
\label{eq:Evir_final}
\end{eqnarray}

Assuming approximate conservation of the effective energy between turnaround and virialization, one obtains
\begin{eqnarray}
W_{{\rm N},t}
+
U_{{\rm EiBI},t}
\simeq
\frac{1}{2}
W_{{\rm N},{\rm vir}}
-
\frac{1}{2}
U_{{\rm EiBI},{\rm vir}} .
\label{eq:virialization_condition_energy}
\end{eqnarray}
This relation provides an implicit estimate of the virial radius once the turnaround configuration and the regularized profile have been specified. Its use should be understood as part of the effective spherical-collapse prescription adopted in this work.

\subsection{Effective top-hat estimate and the Einstein--de Sitter benchmark}

Although EiBI gravity does not admit an exactly discontinuous top-hat profile when the gradient terms are retained, it is still informative to obtain a simple analytic estimate by approximating the interior region as nearly uniform. This approximation should be regarded as an effective interior description of a smooth profile, not as a literal discontinuous top-hat limit.

For a configuration of total mass $M$ and characteristic physical radius $R$, we write
\begin{eqnarray}
\rho
\simeq
\frac{3M}{4\pi R^3}.
\label{eq:rho_uniform_effective}
\end{eqnarray}
The Newtonian contribution is then approximated by
\begin{eqnarray}
W_{\rm N}
\simeq
-\frac{3}{5}
\frac{G_{\rm N}M^2}{R},
\label{eq:WN_uniform}
\end{eqnarray}
while the EiBI bulk contribution becomes
\begin{eqnarray}
U_{\rm EiBI}
\simeq
\frac{\kappa_{\rm BI}}{8}
\rho^2
\left(
\frac{4\pi R^3}{3}
\right)
=
\frac{3\kappa_{\rm BI}}{32\pi}
\frac{M^2}{R^3}.
\label{eq:Ueibi_uniform}
\end{eqnarray}
Substituting Eqs.~\eqref{eq:WN_uniform} and \eqref{eq:Ueibi_uniform} into Eq.~\eqref{eq:virialization_condition_energy}, we obtain the effective relation
\begin{eqnarray}
-\frac{3}{5}
\frac{G_{\rm N}M^2}{R_t}
+
\frac{3\kappa_{\rm BI}}{32\pi}
\frac{M^2}{R_t^3}
\simeq
-\frac{3}{10}
\frac{G_{\rm N}M^2}{R_{\rm vir}}
-
\frac{3\kappa_{\rm BI}}{64\pi}
\frac{M^2}{R_{\rm vir}^3}.
\label{eq:Rvir_equation_effective}
\end{eqnarray}
For $\kappa_{\rm BI}=0$, this equation reduces to the Einstein--de Sitter benchmark
\begin{eqnarray}
R_{\rm vir}
=
\frac{R_t}{2}.
\label{eq:Rvir_half_Rt}
\end{eqnarray}
For nonzero $\kappa_{\rm BI}$, the ratio $R_{\rm vir}/R_t$ is no longer expected to be universal within this effective treatment. It depends on the EiBI coupling, on the enclosed mass, and on the profile assumptions used to approximate the interior and boundary structure of the collapsing region.

\subsection{Virial overdensity as a profile-dependent effective observable}

Once the virial radius has been estimated, the virial overdensity is defined as
\begin{eqnarray}
\Delta_{\rm vir}(a_{\rm vir})
\equiv
\frac{\rho_{\rm th}(a_{\rm vir})}
{\rho_{\rm crit}(a_{\rm vir})}
=
\frac{
\bar\rho_m(a_{\rm vir})
\left[
1+\delta_{\rm th}(a_{\rm vir})
\right]
}
{\rho_{\rm crit}(a_{\rm vir})}
=
\Omega_m(a_{\rm vir})
\left[
1+\delta_{\rm th}(a_{\rm vir})
\right],
\label{eq:Delta_vir_new_def}
\end{eqnarray}
where
\begin{eqnarray}
\rho_{\rm crit}(a)
=
\frac{3H^2(a)}{8\pi G_{\rm N}}.
\label{eq:rho_crit_new_def}
\end{eqnarray}
If the interior remains approximately homogeneous, one may further use
\begin{eqnarray}
1+\delta_{\rm th}(a)
\simeq
\frac{
1+\delta_{\rm th}(a_i)
}
{y(a)^3},
\label{eq:delta_th_effective}
\end{eqnarray}
where $y(a)$ is the dimensionless radius variable introduced in the turnaround analysis.

Thus, $\Delta_{\rm vir}$ remains a useful late-time collapse diagnostic, but in EiBI gravity it should be interpreted as an effective and profile-dependent quantity. This is not simply a technical complication. It reflects the fact that the same matter-gradient correction that affects the collapse equation also enters the virial balance through bulk and surface terms. Therefore, the virial overdensity provides a natural observable through which the profile sensitivity of EiBI gravity may become manifest.

\subsection{Scope of the effective virial prescription}

The derivation presented above has a limited and explicitly effective scope. First, it assumes a smooth regularized density profile. This assumption is essential in EiBI gravity, because the force law contains explicit density gradients. A discontinuous top-hat profile would generate singular boundary terms and is not an admissible configuration once the EiBI correction is retained.

Second, the virial manipulation naturally generates the surface term $\Sigma_{\rm EiBI}$. This term vanishes only under additional assumptions about the behavior of the density at the effective boundary. For a realistic smooth transition layer it need not be exactly zero, and its quantitative importance depends on the coarse-graining prescription. Neglecting $\Sigma_{\rm EiBI}$ should therefore be understood as a leading approximation.

Third, the reduction to a single characteristic radius $R$ is not an exact description of the full EiBI collapse problem. In general, the gradient correction couples the dynamics to the radial structure of the density field. A more complete treatment of virialization would require following the evolution of the full regularized profile and its boundary layer.

For these reasons, the result derived in this section should be regarded as an effective and physically motivated prescription for regularized spherical collapse in EiBI gravity. Its importance lies in making explicit how the EiBI matter-gradient coupling modifies the usual virial balance and makes the virial overdensity sensitive to the density profile. In the numerical analysis we use this prescription as a controlled phenomenological estimate: we compute the turnaround configuration, solve Eq.~\eqref{eq:Rvir_equation_effective} for $R_{\rm vir}$ under the chosen regularization, and then evaluate $\Delta_{\rm vir}$ from Eq.~\eqref{eq:Delta_vir_new_def}. This provides a direct comparison with the $\Lambda$CDM reference solution while keeping the profile dependence of the virialization stage explicit.

\section{Some numerical results}
\label{sec:numerical_results}

We now discuss the numerical behavior of the collapse observables introduced in the previous sections. The purpose of this section is to isolate how the EiBI matter-gradient coupling modifies the standard $\Lambda$CDM spherical-collapse benchmarks once a regularized density profile and an effective coarse-graining prescription have been specified. Since the EiBI correction depends explicitly on spatial derivatives of the density field, a meaningful comparison between different profiles requires first removing trivial differences in normalization, characteristic scale, and enclosed mass. For this reason, the Tanh and peak-based profiles are calibrated to have the same half-amplitude radius and the same cumulative mass proxy inside a common matching region.
The factor $\Omega_m(a)/a^2$ is the factor associated with the effective physical-gradient closure adopted in the numerical implementation.

Figure 1 shows the result of this matching procedure. The left panel compares the normalized radial profiles, while the right panel shows the corresponding cumulative mass proxy. The two profiles are constructed to agree in their global scale and integrated matter content, but they still retain different internal shapes and boundary-layer structures. This residual difference is the relevant one in EiBI gravity, because the weak-field correction is sourced by density gradients. Therefore, after the matching, differences between the Tanh and peak-based collapse observables can be interpreted as genuine profile-shape effects rather than as consequences of using perturbations with different mass normalization.

We first consider the linear collapse threshold. The upper block in Figure 2 shows $\delta_c(z_{\rm coll})$ together with its relative deviation with respect to the $\Lambda$CDM prediction. In the EiBI cases, $\delta_c$ is reduced relative to the reference model, and the deviation increases with the dimensionless coupling $\hat\kappa_{\rm BI}$ over the range considered. The effect remains moderate, but the ordering with the coupling is clear. At fixed coupling, the Tanh and peak-based profiles do not give identical thresholds, showing that even the linearly extrapolated collapse benchmark retains a controlled dependence on the internal structure of the regularized overdensity. Physically, the reduction of $\delta_c$ means that collapse is reached from a slightly smaller linear amplitude once the EiBI matter-gradient contribution is included.

The turnaround overdensity is displayed in the lower block in Figure 2. In contrast with the behavior of $\delta_c$, the quantity $\delta_t(z_{\rm coll})$ is enhanced with respect to $\Lambda$CDM. The enhancement is again ordered by the EiBI coupling and shows a mild but persistent separation between the Tanh and peak-based profiles. This indicates that the EiBI correction does not simply shift the collapse threshold by an approximately uniform amount. Instead, it changes the nonlinear trajectory of the perturbation: the configuration reaches maximum expansion at a larger overdensity than in the standard case. In this sense, $\delta_t$ is a more sensitive nonlinear diagnostic of the EiBI gradient correction than $\delta_c$.

Figure 3 presents the turnaround radius $R_t$ as a function of the collapse redshift. The EiBI correction produces a reduction of $R_t$ relative to the $\Lambda$CDM reference solution, although this effect is weaker than the corresponding changes in the overdensity variables. This hierarchy is physically natural. Since the EiBI modification is sourced by spatial derivatives of the matter density, its impact is expected to be more directly visible in density-based quantities than in the characteristic size of the collapsing region. The turnaround radius therefore provides a complementary geometric diagnostic of the same modified nonlinear evolution.

The mass dependence of the turnaround sector is shown in Figure 4. In the $\Lambda$CDM reference case, the corresponding turnaround observable is nearly universal, or at least only weakly mass dependent, over the interval considered. In EiBI gravity, by contrast, the curves develop a visible mass dependence. This is a direct consequence of the matter-gradient nature of the theory. For a fixed profile prescription, changing the mass changes the characteristic size of the perturbation and therefore the effective strength of the gradient contribution. As a result, objects with different masses do not collapse in a strictly homologous way. The emergence of this mass dependence is one of the clearest indications that EiBI is not equivalent to a simple rescaling of the Newtonian force.

Figure 5 illustrates the evolution of the normalized radius variable used to identify the turnaround condition. This plot provides a direct dynamical view of how the EiBI correction affects the expansion, turnaround, and subsequent contraction of the collapsing shell. The deviations from the $\Lambda$CDM trajectory remain controlled, but the shift in the turnaround epoch and radius is consistent with the behavior inferred from $R_t(z_{\rm coll})$ and $\delta_t(z_{\rm coll})$. This confirms that the changes observed in the turnaround quantities arise from the modified nonlinear evolution rather than from a purely algebraic redefinition of the observables.

We now discuss the virial overdensity, shown in Figure 6, which was obtained from the effective EiBI virial prescription. The quantity $\Delta_{\rm vir}$ is enhanced relative to the $\Lambda$CDM prediction, with a larger deviation for larger $\hat\kappa_{\rm BI}$ and a residual separation between the Tanh and peak-based profiles. This result is consistent with the virial discussion of Sec.~\ref{sec:virial_eibi}. Since the EiBI correction contributes to the effective energy balance through density-dependent bulk terms and, in general, through boundary terms, the final virial overdensity is not expected to remain universal in the same sense as in the general-relativistic top-hat model. The enhancement of $\Delta_{\rm vir}$ therefore provides a late-time counterpart to the enhancement of $\delta_t$.

Figure~7 shows the virial-to-turnaround radius ratio
$\eta\equiv R_{\rm vir}/R_t$ as a function of the collapse redshift. In the $\Lambda$CDM reference case the effective prescription gives the standard value $\eta_{\rm LCDM}=1/2$, represented by the nearly horizontal blue curve. The EiBI corrections produce small positive deviations from this value, with $\eta$ remaining very close to $0.5$ over the whole range $0\leq z_{\rm coll}\leq 3$. The departure increases monotonically with redshift and is larger for $\kappa=10^{-6}$ than for $\kappa=3\times10^{-7}$. For a fixed value of
$\kappa$, the Tanh regularization gives a slightly larger shift than the
peak-based profile. The lower panel displays $\eta-\eta_{\rm LCDM}$, making explicit that the modified-virialization effect is at the level of
$10^{-6}$--$10^{-5}$ for the parameters shown, while preserving the same
hierarchy between $\kappa$ and profile choices observed in the upper panel.

Taken together, the numerical results define a coherent pattern. Relative to $\Lambda$CDM, the EiBI correction lowers the linear collapse threshold $\delta_c$, raises the turnaround overdensity $\delta_t$ and the virial overdensity $\Delta_{\rm vir}$, and produces a more modest reduction of the turnaround radius $R_t$. The deviations are monotonic in the coupling strength over the parameter range considered. Equally important, the Tanh and peak-based profiles do not lead to identical predictions even after matching their characteristic radius and cumulative mass proxy. This confirms that the profile dependence is not a normalization artefact, but a genuine consequence of the gradient-sensitive structure of EiBI gravity.

The physical interpretation is therefore straightforward. EiBI gravity differs from $\Lambda$CDM not by introducing a universal long-range force in vacuum, but by activating a local matter-curvature response controlled by the spatial structure of the density field. As a result, observables that probe the nonlinear density state of the collapsing configuration, especially $\delta_t$ and $\Delta_{\rm vir}$, are more sensitive to the EiBI correction than radius-based quantities. The present results support the use of spherical collapse as a controlled framework for identifying the imprint of EiBI matter-gradient couplings and for preparing future applications to halo abundances and nonlinear structure-formation statistics.

\begin{figure}[h!]
    \includegraphics[width=0.400\textwidth]{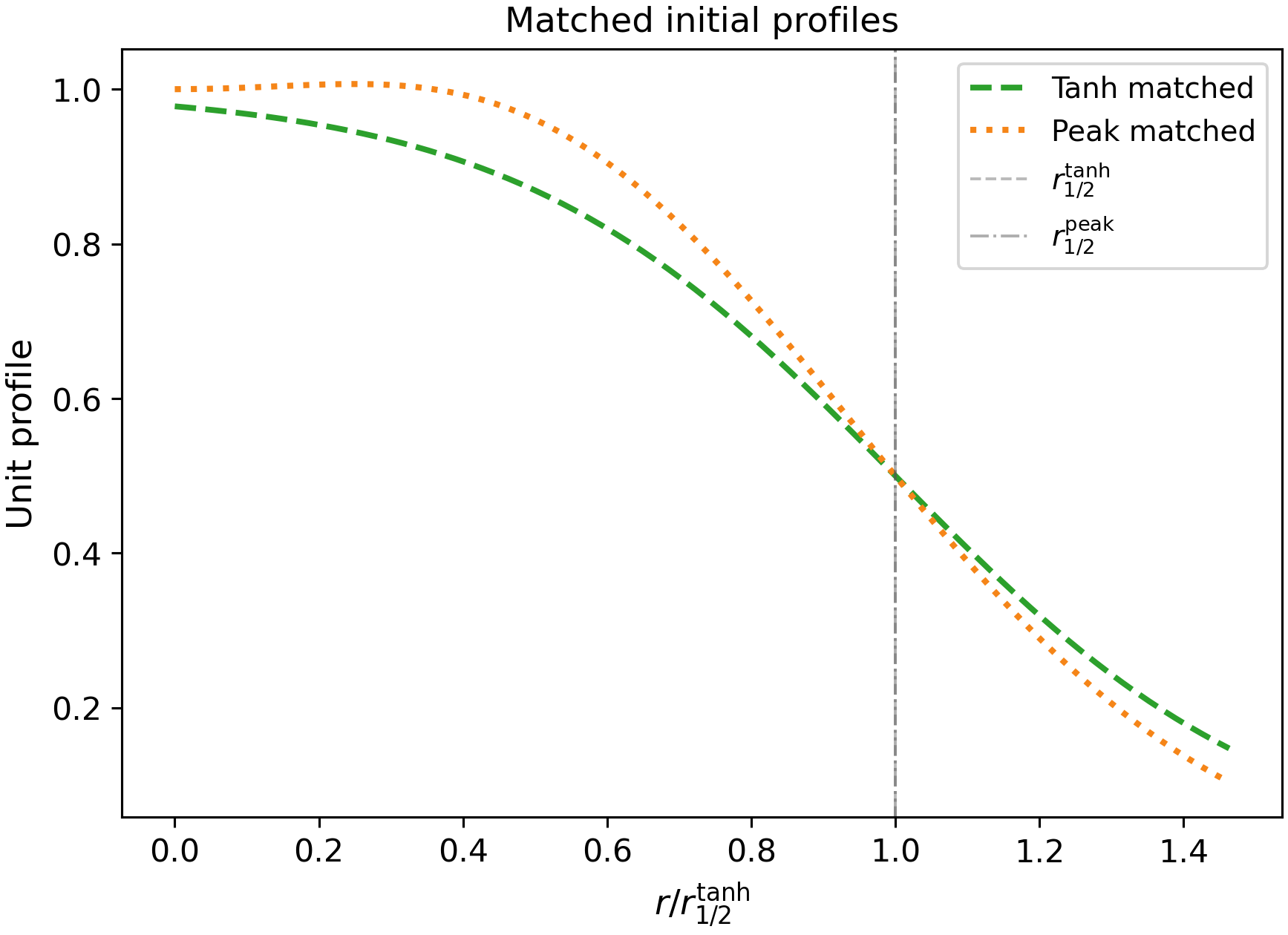}
    \includegraphics[width=0.400\textwidth]{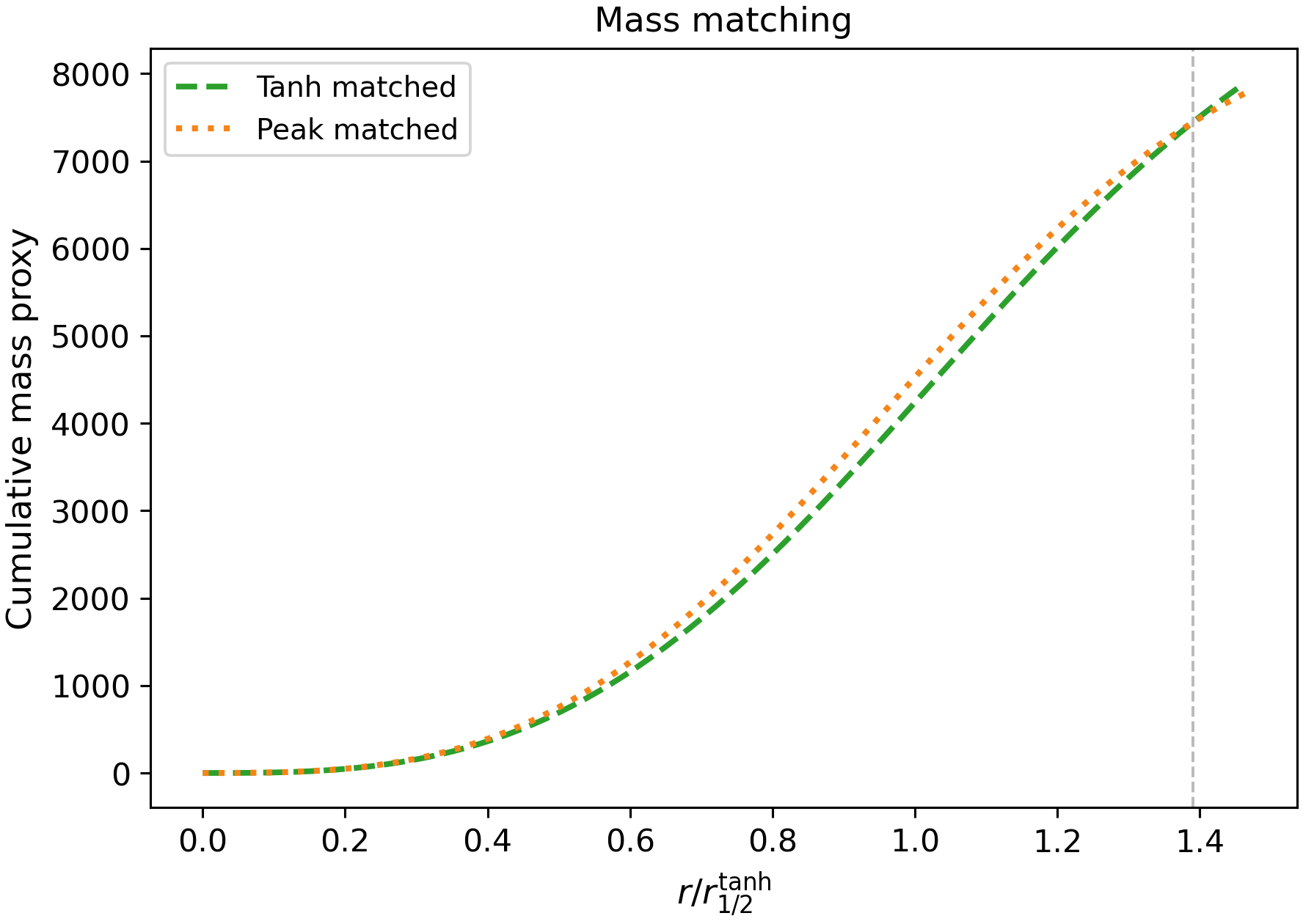}
  \caption{
Matched initial profiles and cumulative mass proxy. 
The peak-based profile is computed with $\nu=2.2$, $n_s=0.965$, a BBKS transfer function with $\Gamma=0.2$, and $R=r_L$ for $M=10^{14}M_\odot/h$. 
The Tanh profile is matched by fixing the half-amplitude radius and the cumulative mass proxy.
}
\label{fig:matched_profiles}
    \label{fig:roche}
\end{figure}

\begin{figure}[h!]
    \includegraphics[width=0.6500\textwidth]{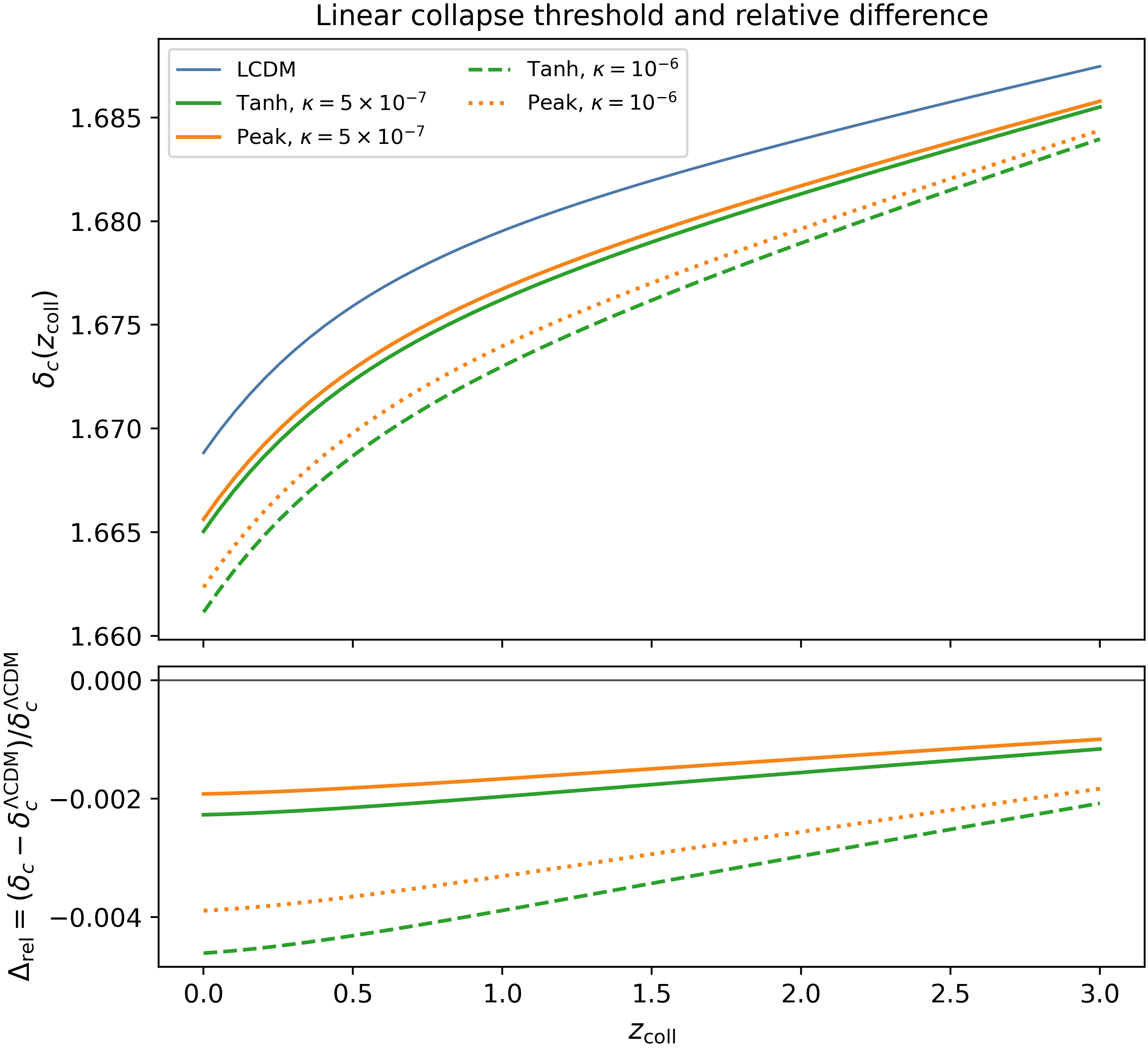}  
    \includegraphics[width=0.650\textwidth]{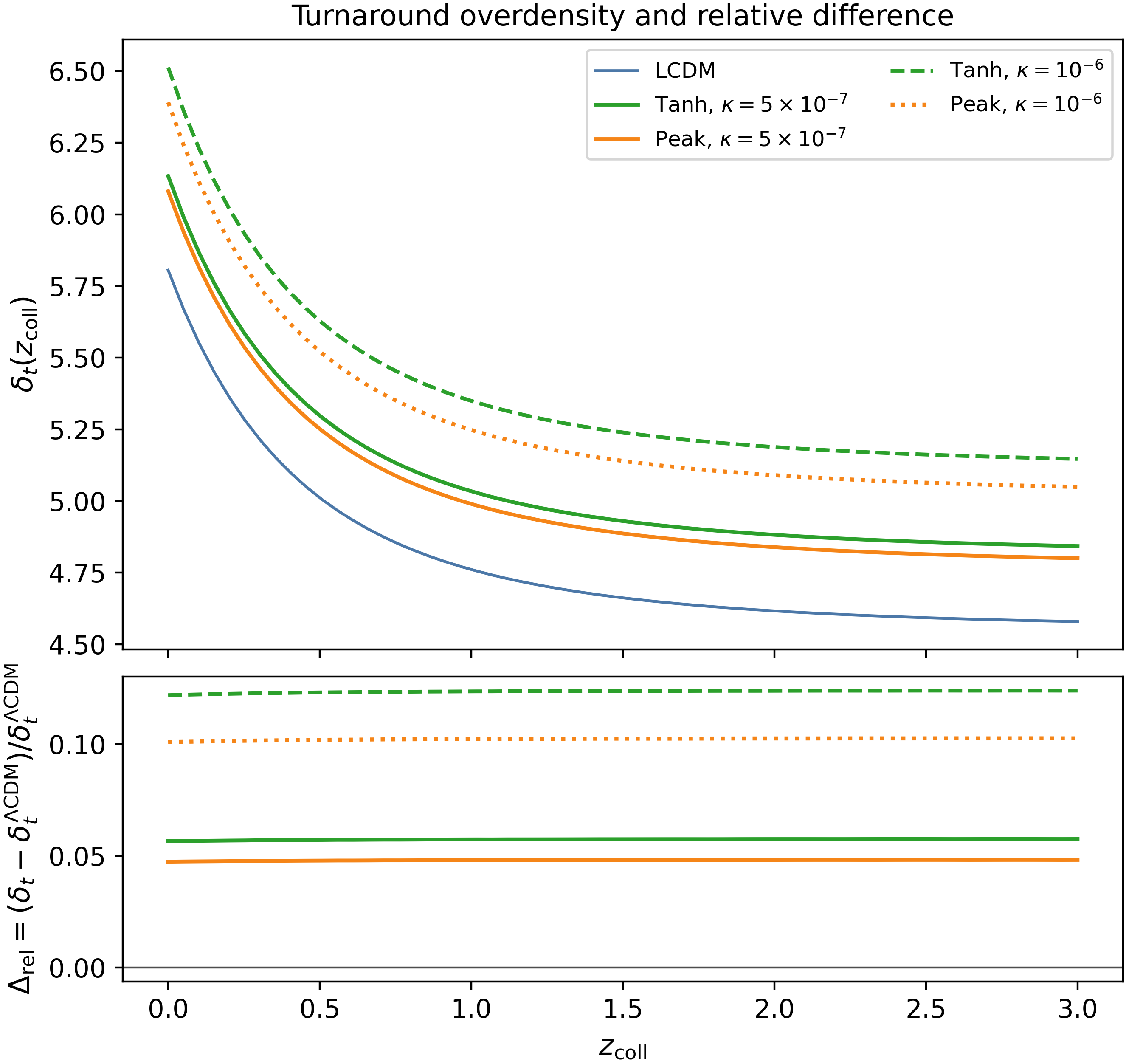}
  \caption{
Linear collapse threshold and turnaround overdensity in EiBI gravity. 
The upper block shows $\delta_c(z_{\rm coll})$ and its relative deviation from $\Lambda$CDM, while the lower block shows $\delta_t(z_{\rm coll})$ and its relative deviation. 
The fiducial parameters are $\Omega_{m0}=0.315$, $\Omega_{\Lambda0}=0.685$, $h=0.674$, $H_0=67.4\,{\rm km\,s^{-1}Mpc^{-1}}$, $a_i=10^{-3}$, $\alpha_k=1$, and $M=10^{14}M_\odot/h$. 
EiBI curves use matched Tanh and peak-based profiles with $\hat\kappa_{\rm BI}=5\times10^{-7}$ and $10^{-6}$.
}
\label{fig:delta_c_delta_t}
\end{figure}

\begin{figure}[h!]
    \includegraphics[width=0.650\textwidth]{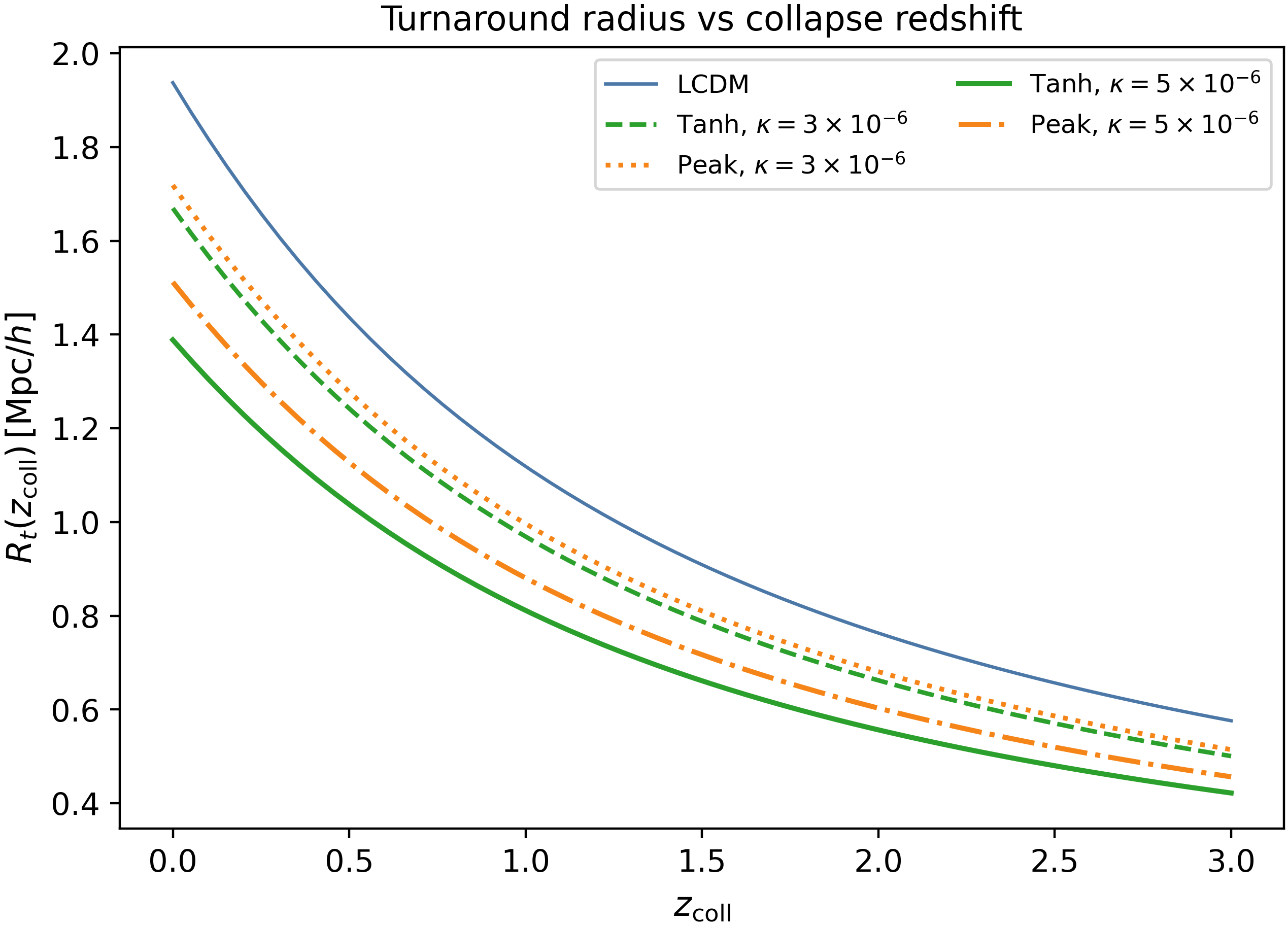}
   \caption{
Turnaround radius $R_t(z_{\rm coll})$ for $\Lambda$CDM and EiBI gravity. 
The calculation uses $\Omega_{m0}=0.315$, $\Omega_{\Lambda0}=0.685$, $h=0.674$, $H_0=67.4\,{\rm km\,s^{-1}Mpc^{-1}}$, $a_i=10^{-3}$, $\alpha_k=1$, and $M=10^{14}M_\odot/h$. 
EiBI curves correspond to matched Tanh and peak-based profiles for the values of $\hat\kappa_{\rm BI}$ shown in the figure.
}
\label{fig:turnaround_radius_zcoll}
\end{figure}

\begin{figure}[h!]
    \includegraphics[width=0.650\textwidth]{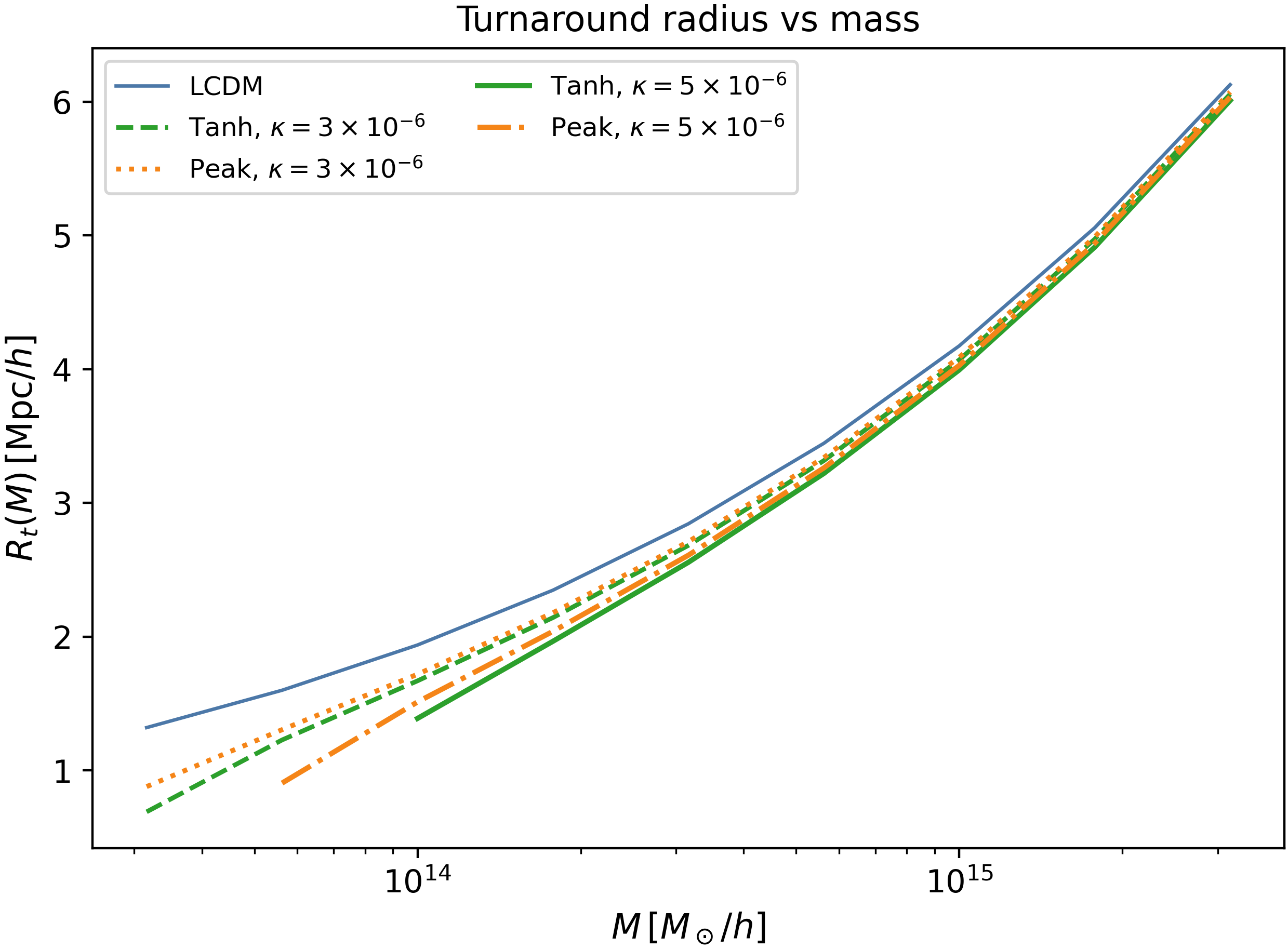}
  \caption{
Mass dependence of the turnaround radius $R_t(M)$. 
The background is fixed by $\Omega_{m0}=0.315$, $\Omega_{\Lambda0}=0.685$, $h=0.674$, and $H_0=67.4\,{\rm km\,s^{-1}Mpc^{-1}}$, with $a_i=10^{-3}$ and $\alpha_k=1$. 
EiBI curves compare matched Tanh and peak-based profiles for the coupling values indicated in the figure.
}
\label{fig:turnaround_radius_mass}
\end{figure}

\begin{figure}[h!]
    \includegraphics[width=0.650\textwidth]{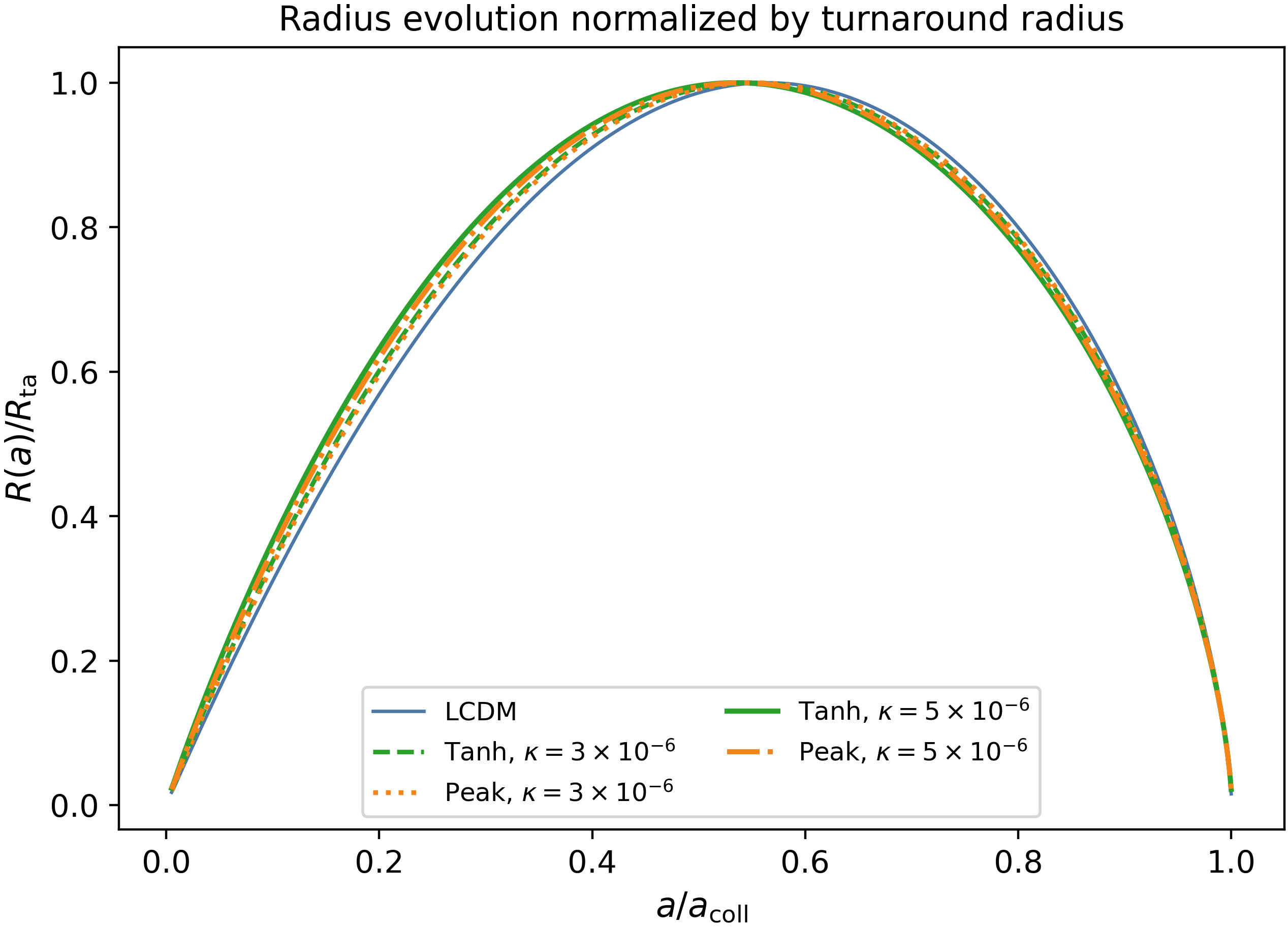}
   \caption{
Physical-radius evolution normalized by the turnaround radius, $R(a)/R_t$, as a function of $a/a_{\rm coll}$. 
The maximum $R(a)/R_t=1$ identifies the turnaround epoch. 
The calculation uses $\Omega_{m0}=0.315$, $\Omega_{\Lambda0}=0.685$, $h=0.674$, $H_0=67.4\,{\rm km\,s^{-1}Mpc^{-1}}$, $a_i=10^{-3}$, $\alpha_k=1$, and $M=10^{14}M_\odot/h$. 
EiBI curves use matched Tanh and peak-based profiles for the coupling values shown in the figure.
}
\label{fig:radius_evolution}
\end{figure}

\begin{figure}[h!]
    \includegraphics[width=0.650\textwidth]{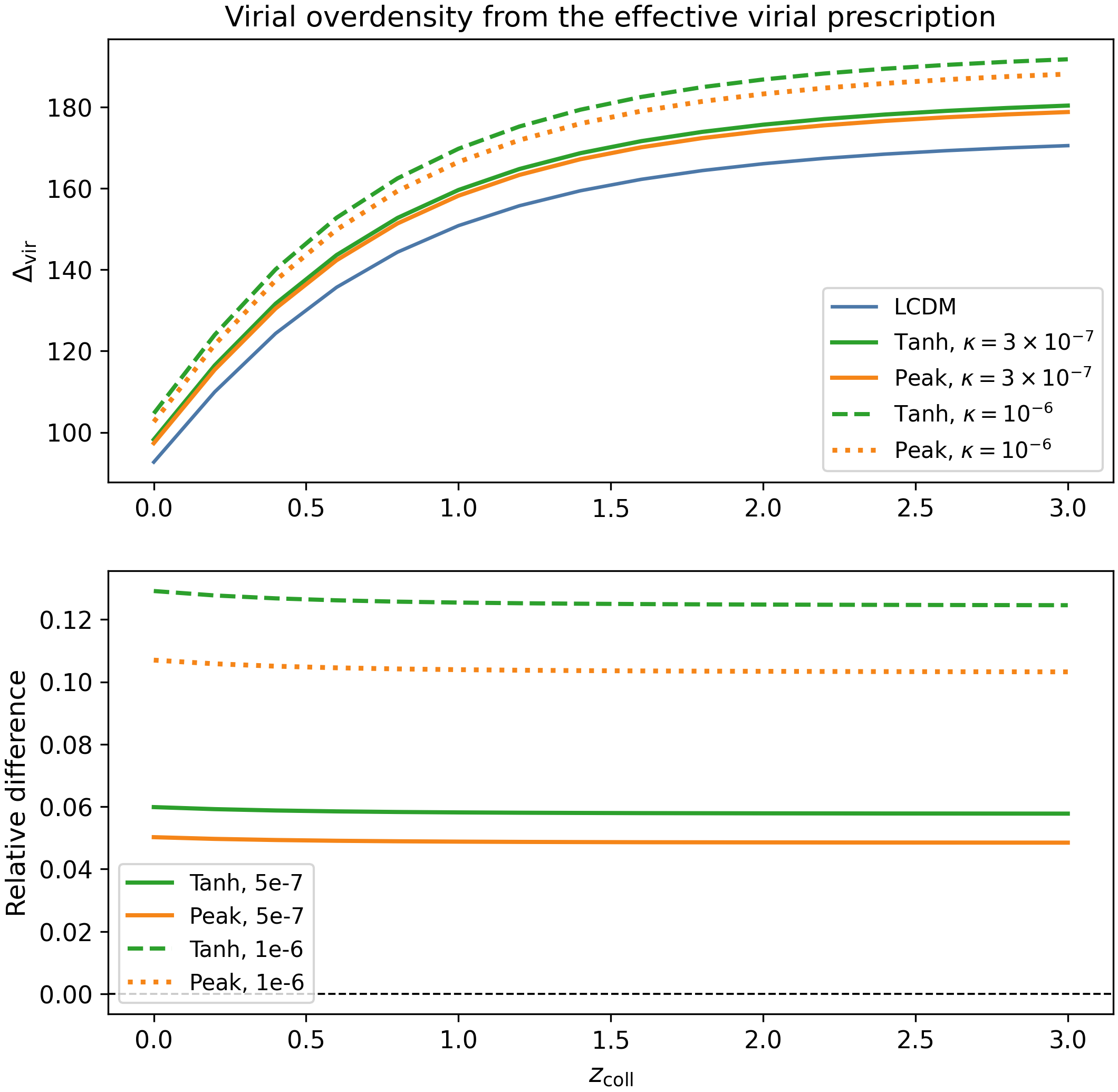}
  \caption{
Virial overdensity $\Delta_{\rm vir}(z_{\rm coll})$ obtained from the effective EiBI virial prescription. 
The calculation adopts $\Omega_{m0}=0.315$, $\Omega_{\Lambda0}=0.685$, $h=0.674$, $H_0=67.4\,{\rm km\,s^{-1}Mpc^{-1}}$, $a_i=10^{-3}$, $\alpha_k=1$, and $M=10^{14}M_\odot/h$. 
EiBI curves compare matched Tanh and peak-based profiles for the values of $\hat\kappa_{\rm BI}$ indicated in the figure.
}
\label{fig:virial_overdensity}
\end{figure}

\begin{figure}[h!]
    \includegraphics[width=0.650\textwidth]{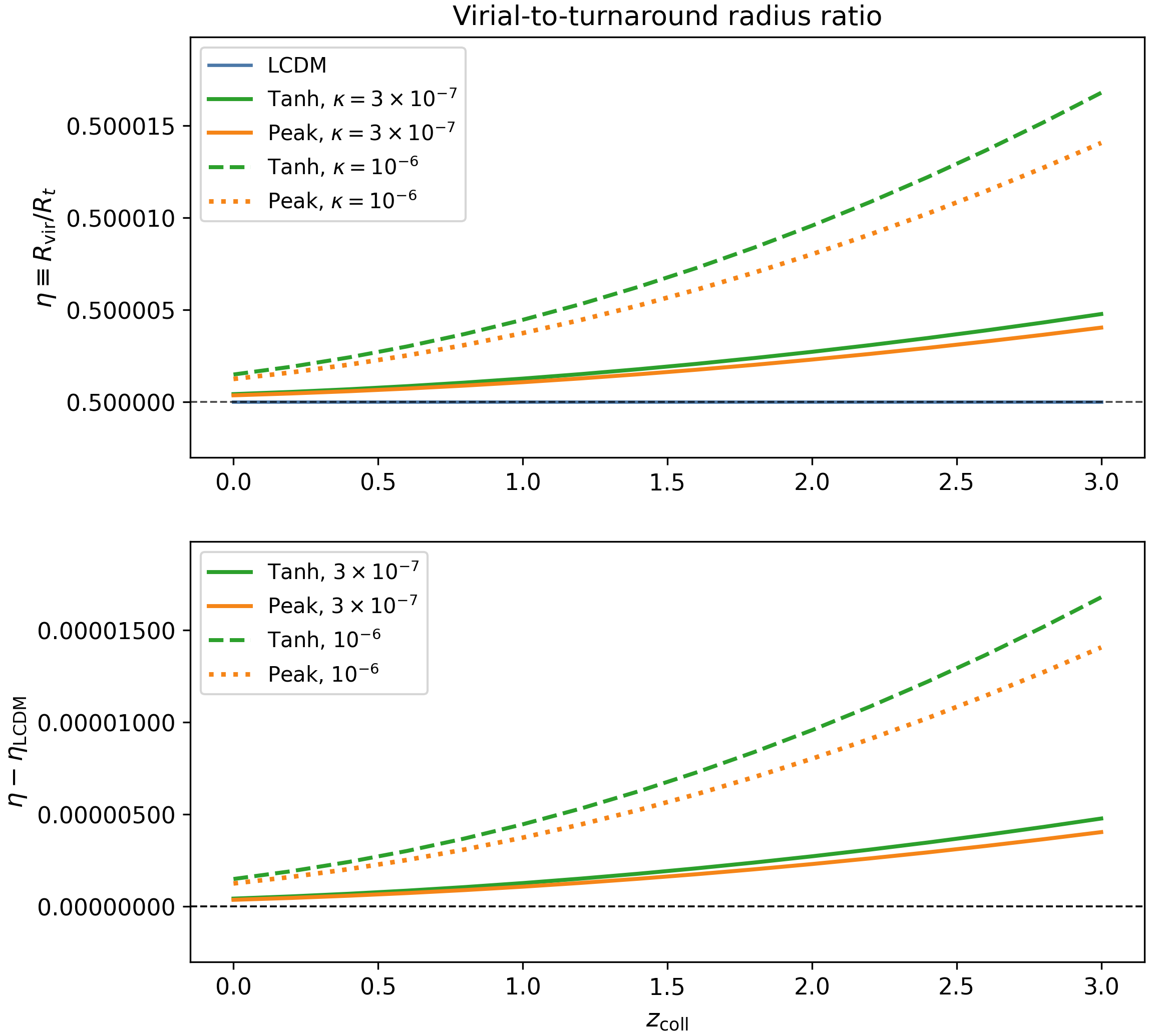}
  \caption{
Virial-to-turnaround radius ratio $\eta\equiv R_{\rm vir}/R_t$ as a function of the collapse redshift. The LCDM reference case gives the standard effective value $\eta_{\rm LCDM}=1/2$, shown by the nearly constant blue curve. The EiBI correction produces small but systematic departures from this value, enhanced at larger $z_{\rm coll}$ and for larger $\kappa$. The lower panel shows $\eta-\eta_{\rm LCDM}$, isolating the modified-virialization contribution and the residual dependence on the regularized density profile.
}
\end{figure}

\section{Conclusions}
\label{sec:conclusions}

In this work we have developed a spherical-collapse framework for Eddington-inspired Born--Infeld gravity in the subhorizon, pressureless, and quasi-static regime. The central point of the analysis is that the weak-field limit of EiBI gravity contains a matter-gradient correction, so that the collapse dynamics is not determined only by the overdensity amplitude, as in the standard general-relativistic top-hat picture. Instead, the spatial structure of the matter distribution becomes part of the physical problem. For this reason, an ideal discontinuous top-hat profile is not an admissible starting point once the EiBI correction is retained, since it would generate singular contributions at the boundary. A regularized density profile and an associated coarse-graining prescription are therefore not merely numerical devices, but intrinsic ingredients of spherical collapse in EiBI gravity.

Starting from the nonlinear continuity and Euler equations, we derived the corresponding evolution equation for the density contrast and made explicit how the EiBI contribution enters through spatial derivatives of the density field. We also clarified the relation between comoving and physical Laplacians and adopted, for the numerical implementation, an effective physical-gradient closure for the EiBI source term. This prescription provides a controlled way of encoding the profile dependence through a dimensionless factor while keeping the calculation suitable for comparison with the standard $\Lambda$CDM spherical-collapse benchmarks. Within this framework, we compared two regularized initial configurations: a Tanh profile and a peak-based profile. These profiles were matched in characteristic radius and cumulative mass proxy, so that the remaining differences between their predictions can be interpreted as genuine profile-shape effects rather than trivial normalization effects.

The numerical results exhibit a coherent pattern. Relative to the $\Lambda$CDM reference model, the EiBI correction lowers the linear collapse threshold $\delta_c$, enhances the turnaround overdensity $\delta_t$, and increases the virial overdensity $\Delta_{\rm vir}$. The turnaround radius $R_t$ is instead moderately reduced, with a weaker response than the overdensity variables. The deviations are ordered by the dimensionless coupling $\hat\kappa_{\rm BI}$ over the parameter range explored. This hierarchy indicates that density-based quantities are more sensitive to the EiBI matter-gradient correction than radius-based observables. In particular, the turnaround and virial overdensities appear as especially useful nonlinear diagnostics of the Born--Infeld character of the theory.

An important outcome of the analysis is the residual dependence on the internal shape of the regularized profile. Even after matching the Tanh and peak-based configurations in radius and cumulative mass proxy, their collapse observables do not coincide exactly. This shows that the profile dependence is not an artefact of normalization, but a physical consequence of the gradient-sensitive EiBI force. In addition, the turnaround sector develops a visible mass dependence, in contrast with the near-universality of the standard top-hat treatment. This result reinforces the interpretation that EiBI gravity does not act as a simple rescaling of Newton's constant. Rather, it introduces a local matter-curvature response controlled by the characteristic size and radial structure of the collapsing object.

We also introduced an effective virial prescription adapted to regularized spherical collapse in EiBI gravity. This construction should not be interpreted as a universal virial theorem for arbitrary EiBI configurations. Its role is more limited and more specific: it provides a physically motivated estimate of the virial radius and virial overdensity once a smooth coarse-grained profile has been specified. The weak-field EiBI correction generates a bulk contribution proportional to $\int\rho^2 dV$ and, in general, a surface term associated with the boundary structure of the density profile. This makes the virial overdensity an effective, profile-dependent quantity rather than a universal function of redshift alone. In the limit $\kappa_{\rm BI}\rightarrow0$, the standard general-relativistic virial relation and the usual Einstein--de Sitter benchmark are recovered, providing a useful consistency check of the prescription.

The present work should therefore be regarded as a first step toward a more complete phenomenological treatment of nonlinear structure formation in EiBI gravity. Its main purpose has been to establish the collapse sector in a controlled way: to derive the relevant equations, clarify the role of regularization, quantify the impact of the matter-gradient correction on $\delta_c$, $\delta_t$, $R_t$, and $\Delta_{\rm vir}$, and identify which quantities are most sensitive to the EiBI modification. This step is necessary before using the theory to make predictions for halo abundances or other large-scale-structure observables, since these applications depend directly on the collapse threshold, the virial overdensity, and their mass and redshift dependence.

Several extensions naturally follow from this analysis. First, the dependence on the smoothing prescription should be explored more systematically by considering broader families of regularized profiles and by quantifying the sensitivity of the results to the coarse-graining scale. Second, the effective virial prescription can be refined by treating the surface contribution more explicitly and by following the full radial evolution of the regularized profile beyond the single-radius approximation. Third, the collapse observables derived here can be incorporated into semi-analytic descriptions of halo formation, including halo mass functions, halo bias, cluster abundances, and number counts. Such an extension would allow one to connect the EiBI matter-gradient coupling with potentially observable signatures in nonlinear large-scale structure.

In this sense, the present paper provides the theoretical basis for a subsequent phenomenological program. Having established how EiBI gravity modifies the spherical-collapse benchmarks, the next step will be to propagate these modifications into halo statistics and number-count predictions. This will make it possible to assess whether the profile and mass dependence found here can lead to distinguishable signatures in cluster surveys and other probes of nonlinear cosmic structure. A full confrontation with observational data is beyond the scope of the present work, but the results obtained here define the necessary ingredients for carrying out that analysis in future studies.

\section*{Acknowledgments}

I would like to thank Maria Margarita and Miguel Amado for their continuous inspiration and support, and the Foundation for Research Support of Espírito Santo (FAPES) for the partial support for the present work.

\appendix

\section{Explicit form of the kernel for the physical initial profile}
\label{app:phy_kernel}

For completeness, we present in this appendix the explicit expression of the
kernel $F(\nu,n_s,k,R)$ that enters the primordial shape function used to
construct the peak-based, or physical, initial profile. This profile follows
the prescription adopted in Refs.~\cite{Kopp2013,BBKS1986}, where the initial
overdensity is interpreted as the mean density contrast around statistically
selected peaks of a Gaussian random field. In this approach, the dependence on
the peak height, on the primordial spectral index, and on the smoothing scale
is encoded in the kernel $F(\nu,n_s,k,R)$.

The explicit form of this kernel is
\begin{eqnarray}
F(\nu,n_s,k,R)
&=&
\left[
\frac{
e^{-\frac{1}{8}\left(\frac{n_s+3}{n_s+5}\right)^{3/2}\nu^2}
\left[
(12n_s+60)\,
e^{\frac{1}{8}\left(\frac{n_s+3}{n_s+5}\right)^{3/2}\nu^2}
+
(0.632\,n_s+13.52)n_s+44.6
\right]
}{
(n_s+5)^2
\left[
2\sqrt{(0.25\,n_s+0.75)\nu^2+0.45\,n_s+8.25}
+
\sqrt{\frac{n_s+3}{n_s+5}}\,\nu
\right]
}
\right]
\nonumber\\[0.2cm]
&&\times
\left[
\left(
\sqrt{\frac{n_s+3}{n_s+5}}\,
\nu\,
\Gamma\!\left(\frac{n_s+5}{2}\right)
\right)
\left(
2k^2R^2-n_s-3
\right)
\right.
\nonumber\\[0.2cm]
&&\left.
\qquad
+
2(n_s+3)\,
\Gamma\!\left(\frac{n_s+7}{2}\right)
\left(
-2k^2R^2+n_s+3
\right)
+
4(n_s+5)\,
\Gamma\!\left(\frac{n_s+3}{2}\right)
\right] .
\label{eq:F_explicit_appendix}
\end{eqnarray}
Here $\nu$ is the dimensionless height of the selected peak, $n_s$ is the
scalar spectral index, $k$ is the Fourier wavenumber, and $R$ is the smoothing
scale. The symbol $\Gamma(x)$ denotes the Euler gamma function. The first
factor in Eq.~(\ref{eq:F_explicit_appendix}) contains the dependence on the
peak statistics, in particular on $\nu$ and $n_s$, while the second factor
contains the explicit dependence on the dimensionless combination $kR$. Thus,
$F(\nu,n_s,k,R)$ determines how the average profile around a selected peak
differs from a purely phenomenological smoothing prescription.

With this kernel, the primordial shape function can be written as
\begin{eqnarray}
\delta_0(k,R)
&=&
\delta_{i,0}\,
\frac{1}{4\pi (n_s+5) R^3}\,
e^{-k^2R^2}\,
(kR)^{n_s}\,
F(\nu,n_s,k,R) .
\label{eq:delta0_appendix_repeat}
\end{eqnarray}
In this expression, $\delta_{i,0}$ is the central amplitude of the initial
overdensity, $R$ fixes the smoothing scale, and the factor $e^{-k^2R^2}$
suppresses modes with wavelengths much smaller than the smoothing scale. The
factor $(kR)^{n_s}$ carries the primordial spectral dependence, while
$F(\nu,n_s,k,R)$ incorporates the correction associated with conditioning the
initial field on the presence of a peak. The peak height is defined by
\begin{eqnarray}
\nu
&=&
\frac{\delta_{i,0}}{\sigma_i(R)} ,
\label{eq:nu_appendix}
\end{eqnarray}
where $\sigma_i(R)$ is the rms fluctuation of the initial density field
smoothed on the scale $R$.

The corresponding real-space initial profile is obtained through the spherical
Fourier transform
\begin{eqnarray}
\delta_i(r,R)
&=&
\frac{2}{\pi}
\int_0^\infty
dk\,
k^2\,
\delta_0(k,R)\,
\frac{\sin(kr)}{kr}\,
T(k) .
\label{eq:phy_profile_appendix_repeat}
\end{eqnarray}
Here $r$ is the radial coordinate measured from the center of the peak and
$T(k)$ is the matter transfer function. The factor $\sin(kr)/(kr)$ is the
spherical Bessel kernel that appears when transforming an isotropic Fourier
profile into real space. Therefore, Eq.~(\ref{eq:phy_profile_appendix_repeat})
defines the spherically averaged initial overdensity profile associated with a
selected peak of height $\nu$.

This construction is useful for the present work because the EiBI correction
depends explicitly on spatial derivatives of the density field. The collapse
dynamics is therefore sensitive not only to the enclosed mass and the central
amplitude, but also to the detailed radial shape of the initial overdensity.
The peak-based profile provides a statistically motivated alternative to the
regularized Tanh profile, allowing us to test how the EiBI spherical-collapse
observables depend on the internal structure of the initial perturbation while
keeping the same characteristic mass scale.

\end{document}